\documentclass[12pt, letterpaper]{article}
\pdfoutput=1
\usepackage[left=25.4mm,right=25.4mm,top=25.4mm,bottom=25.4mm]{geometry}
\usepackage[utf8]{inputenc}
\usepackage[version=3]{mhchem} 
\usepackage{siunitx}
\usepackage{graphicx}
\usepackage{amsmath}
\usepackage{authblk}
\usepackage{subcaption}
\usepackage{soul}
\usepackage{color}
\usepackage[table]{xcolor}
\begin{document}

\title{Analyzing the dominant SARS-CoV-2 transmission routes towards an \textit{ab initio} SEIR model}
\author[1]{Swetaprovo Chaudhuri \thanks{Corresponding author, email: schaudhuri@utias.utoronto.ca}}
\author[2]{Saptarshi Basu} 
\author[3]{Abhishek Saha} 
\affil[1]{Institute for Aerospace Studies, University of Toronto, Toronto, Canada}
\affil[2]{Department of Mechanical Engineering, Indian Institute of Science, Bengaluru, India}
\affil[3]{Department of Mechanical and Aerospace Engineering, University of California San Diego, La Jolla, USA}

\date{}

\maketitle
\begin{abstract}
Identifying the relative importance of the different transmission routes of the SARS-CoV-2 virus is an urgent research priority. To that end, the different transmission routes, and their role in determining the evolution of the Covid-19 pandemic are analyzed in this work. Probability of infection caused by inhaling virus-laden droplets (initial, ejection diameters between $0.5-750\mu m$) and the corresponding desiccated nuclei that mostly encapsulate the virions post droplet evaporation, are individually calculated. At typical, air-conditioned yet quiescent indoor space, for average viral loading, cough droplets of initial diameter between $10-50 \mu m$ have the highest infection probability. However, by the time they are inhaled, the diameters reduce to about $1/6^{th}$ of their initial diameters. While the initially near unity infection probability due to droplets rapidly decays within the first $25s$, the small yet persistent infection probability of desiccated nuclei decays appreciably only by $\mathcal{O} (1000s)$, assuming the virus sustains equally well within the dried droplet nuclei as in the droplets. Combined with molecular collision theory adapted to calculate frequency of contact between the susceptible population and the droplet/nuclei cloud, infection rate constants are derived \textit{ab initio}, leading to a SEIR model applicable for any respiratory event - vector combination. Viral load, minimum infectious dose, sensitivity of the virus half-life to the phase of its vector and dilution of the respiratory jet/puff by the entraining air are shown to mechanistically determine specific physical modes of transmission and variation in the basic reproduction number $\mathcal{R}_0$, from first principle calculations.
\end{abstract}

\section{Introduction}
One of the longstanding questions of pandemics involving respiratory droplets is identifying their dominant mode of transmission. The most well recognized pathways for contagious respiratory diseases are i) the direct contact/inhalation of the relatively larger infectious droplets ($>5 \mu m$) commonly known as the droplet mode of transmission, ii) airborne or aerosol transmission which is presumed to be caused by inhalation of very small infectious droplets ($<5 \mu m$) floating in air and iii) contact with infectious surfaces - fomites. For the present Covid-19 pandemic, while the droplet mode of transmission is well established, evidence/possibility for aerosol transmission \cite{Morawska2020Commentary, liu2020aerodynamic} renders identifying the dominant transmission route an intriguing scientific problem with extremely high implications for human health and public policy. On July 9, 2020, World Health Organization issued a scientific brief \cite{WHOJuly92020} stating ``Urgent high-quality research is needed to elucidate the relative importance of different transmission routes; the role of airborne transmission in the absence of aerosol generating procedures.'' In this paper, we establish a fundamental theoretical framework where the relative strength of the individual transmission routes are analyzed from first principles with idealizing assumptions. Many biological aspects of the disease transmission, including but not limited to effects of immune response are beyond the scope of this paper and will not be addressed here with exclusive focus on the physical aspects \cite{mittal2020flow} of the disease transmission. Physics is involved in at least four levels in a Covid-19 type pandemic evolution, micro-scale droplet physics, spray/droplet-cloud physics, collision/interaction between the spray/cloud and the susceptible individuals, deposition and absorption of the inhaled droplets/droplet nuclei. First three are addressed in this paper at different levels of complexity. 
We adopt the convention that respiratory droplets (all liquid phase droplets of all sizes, typically 0.5-750$\mu m$) cause disease transmission by ``droplet or $d$'' route, whereas dried or desiccated droplet nuclei which in this paper refers to the semi-solid/crystalline residue that remains after the droplet liquid evaporates, is responsible for the ``dried droplet nuclei or $n$'' route of transmission. Thus, the $d$ route invariably includes droplets less than as well as greater than 5$\mu m$, instead of resorting to the rather arbitrary threshold to distinguish between droplets and nuclei. The reason of our choice is that, the distinct thermodynamic phase of the transmission vector: liquid versus semi-solid is expected to be a much better identifier to delineate the different pathways. Furthermore, the virus survivability within the dried droplet nuclei could be well different from that of the liquid droplet. Small and medium sized droplets do remain airborne after their ejection for substantially long periods of time \cite{stadnytskyi2020airborne} due to the fact that droplet size continuously changes, except in highly humid conditions ($RH_{\infty}>85 \%$), due to evaporation until desiccation. Respiratory droplets are ejected during different expiratory events: breath, cough, sing, sneeze or talk (ing), when the droplets are ejected with different droplet size distributions \cite{duguid1945numbers,xie2009exhaled, chao2009characterization}. In violent expiratory events like coughing or sneezing, the droplets co-move with a turbulent jet of exhaled air. The trajectories of the jet and the droplets could diverge due to aerodynamic drag and gravity effects. Nevertheless, experiments by Bourouiba et al. \cite{bourouiba2014violent, bourouiba2020turbulent} have shown that these droplets can travel rather large distances initially within a turbulent jet which later transitions to a puff or a cloud due to the lack of a continuous momentum source. Depending on the ambient conditions and droplet size, these droplets evaporate at different times. However, while water - the volatile component of the mucosalivary liquid, evaporates, the non-volatile components - salt, protein, mucous and virus particles present, separate out by crystallization processes. These semi-solid droplet residues, typically about 10-20\% of the initial droplet diameter are called dried droplet nuclei, remain floating as aerosols and are believed to be responsible for airborne mode of droplet transmission. While only very few experiments have so far probed the structure of these dried droplet nuclei, the first of its kind work by Vejerano and Marr \cite{vejerano2018physico} provided critical insights on the distribution of virus particles inside the dried droplet nuclei. Marr et al. \cite{marr2019mechanistic} offered mechanistic insights on the role of relative humidity in respiratory droplet evaporation but the question on the role of evaporation, the resulting chemistry inside the dried droplet nuclei on virus survivability persisted. This was explored by Lin and Marr in \cite{lin2019humidity}. It was found that virus survivability inside sessile droplets is a non-monotonic function of ambient relative humidity, and of course dependent on the specific virus type as well. Questions on the survivability of the SARS-CoV-2 virus inside the dried droplet nuclei from contact free droplets, as it would happen for respiratory sprays, is not yet settled. 
In this paper, first we present a model to identify the probability of infection transmission for two different routes $d$ and $n$, by accounting for the corresponding droplet size distribution, viral load and virus half-life. Next, we briefly present the droplet/nuclei cloud aerodynamics and respiratory droplet evaporation physics. $1 \%$ NaCl-water solution is used as a surrogate for the mucosalivary fluid. This is followed by modeling the generalized infection rate constant which can be used in theory for any kind of expiratory event or for any mode of transmission. The rate constant is then incorporated into a SEIR model \cite{keeling2011modeling} using the formalism of a chemical reaction mechanism. Finally, the results and discussions are presented followed by conclusion. 

In most epidemiological models, the rate constants (parameters of the SEIR differential equations) which lead to the $\mathcal{R}_0$ are obtained by fitting available data on the number of new infections \cite{keeling2011modeling, bertozzi2020challenges}.  
Indeed, this type of epidemiological models have provided immense insights on Covid-19 and the necessity of non-pharmaceutical public health interventions \cite{bertozzi2020challenges, adam2020special, metcalf2020mathematical}. 
However, it is to be recognized that the data for a Covid-19 type pandemic is almost always under-reported due to large number of asymptomatic cases. Furthermore, the actual rate constants and $\mathcal{R}_0$ could depend on several physical factors like temperature, relative humidity, UV-index etc. Thus, with changing conditions, parameters obtained from fitting recent past data may not be adequate, standalone, to predict nature of future outbreaks. Therefore, there is a pressing need to develop a framework to understand and calculate $\mathcal{R}_0$ from \textit{ab initio} calculations while being cognizant of the idealizing assumptions and limitations involved in the process. To the authors' knowledge this is the first time aerodynamics and thermodynamics of the droplets/nuclei, viral load, half-life and minimum required viral dose for infection, have been systematically accounted to obtain the rate constants and $\mathcal{R}_0$ of a SEIR model, \textit{ab initio}.

\section{The model}
\subsection{Probability of infection by different transmission routes}
\begin{figure}[htb!]
\includegraphics[width=1\textwidth]{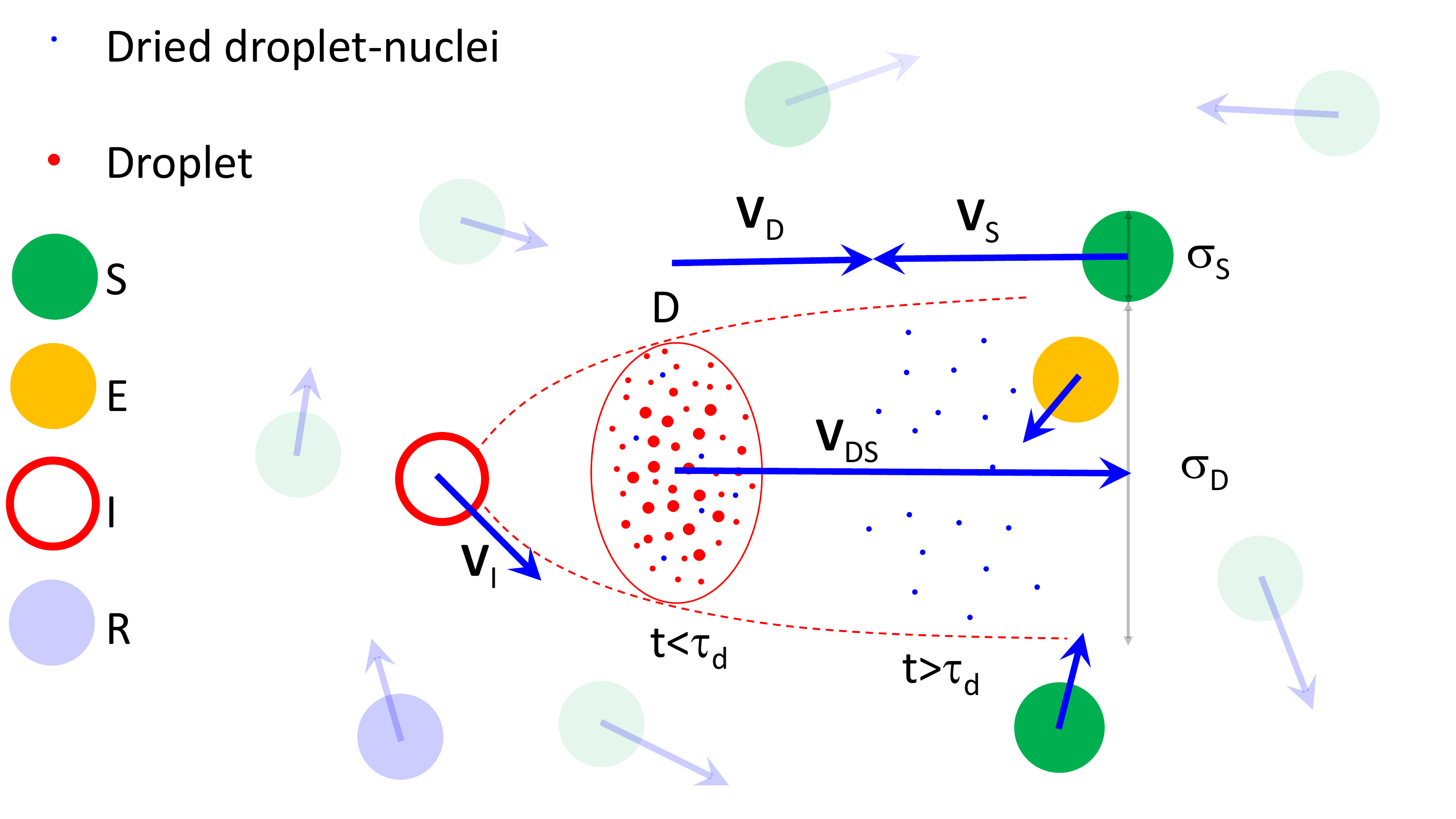}
\begin{centering}
\caption{Schematic of the interaction between $S$ and the droplet/dried droplet nuclei cloud $D$ ejected by $I$, resulting in $E$. $E$ would soon be converted to $I$ which would either result in $R$ or $X$. The red ellipse marks the control volume analyzed for computing the probability of infection $\mathcal{P}_{\alpha\beta}$.}
\label{Fig:Schematic}
\end{centering}
\end{figure}

In this subsection, we estimate the probability of infection by different transmission modes for any expiratory event. Consider an infected person $I$ exhaling a droplet laden jet that quickly transforms into a droplet cloud $D$ in vicinity of susceptible individuals $S$ as shown in Fig. \ref{Fig:Schematic}.

The instantaneous diameter of the respiratory jet/puff/cloud is given by $\sigma_D$, its velocity with respect to $S$ is given by $\vec{V}_{DS}$, while the effective diameter of the hemispherical volume of air inhaled by $S$, for every breath, is given by $\sigma_S$. $\sigma_{DS}=(\sigma_D + \sigma_S)/2$. The primary objective of this subsection is to estimate $\mathcal{P}_{\alpha \beta}(t)$ which denotes the time dependent probability of infection of $S$ for the given expiratory event $\alpha$ and type of transmission vector $\beta$. Thus, $\alpha$ denotes one among breathing, coughing, singing, sneezing or talking, while $\beta$ denotes one among droplets, dried droplet nuclei or fomites. It is to be recognized that $\sigma_{DS}, \vec{V}_{DS}$ are not only functions of time, but also dependent on $\alpha$ and $\beta$, though their subscripts have been dropped for brevity. As the droplet/nuclei cloud entrains surrounding air, it grows in size with concomitant dilution of the particles inside, thereby reducing $\mathcal{P}_{\alpha \beta}$. Of course $\mathcal{P}_{\alpha \beta}$ must also be determined by the viral load and droplet size distribution. $\mathcal{P}_{\alpha \beta}$ could be obtained by solving the transport equations. Instead, here we take a Lagrangian approach of tracking and analyzing the control volume of the air-droplet cloud $D$ ejected by $I$ and droplets/droplet-nuclei within. 

At the moment of the onset of the expiratory event denoted by $t=0$, a log-normal distribution could be used to describe the probability density function (pdf) of the initial droplet size distribution $f_{\alpha}$ of the ejected respiratory spray.
\begin{equation}
f_{\alpha}(D) = \frac{1}{\sqrt{2\pi}\sigma D} e^{-(\ln(D) - \mu)^2/2 \sigma^2}
\end{equation}
$D$ is the sample space variable of the initial droplet diameter $D_{s,0}$. $\mu$ and $\sigma$ are the mean and standard deviation of $ln(D)$. If $N_{t \alpha}$ is the total number of droplets ejected for the expiratory event $\alpha$, the number of droplets within the interval $dD$ is given by
$N_{t \alpha}f_{\alpha}(D)dD$. Therefore, for a given $f_{\alpha}$ and $\rho_v$ - the viral load in number of copies per unit volume of the ejected liquid, the cumulative number of virions in droplets between sizes $D_1$ and $D_2$ is given by 
\begin{equation}
N_{v \alpha}=\frac{\pi \rho_v N_{t \alpha}}{6}\int_{D_1}^{D_2} D^3f_{\alpha}(D)dD
\label{Eq: Nvd}
\end{equation}

\begin{figure}[ht!]
\includegraphics[trim=1cm 0cm 1cm 0cm,clip,width=1\textwidth]{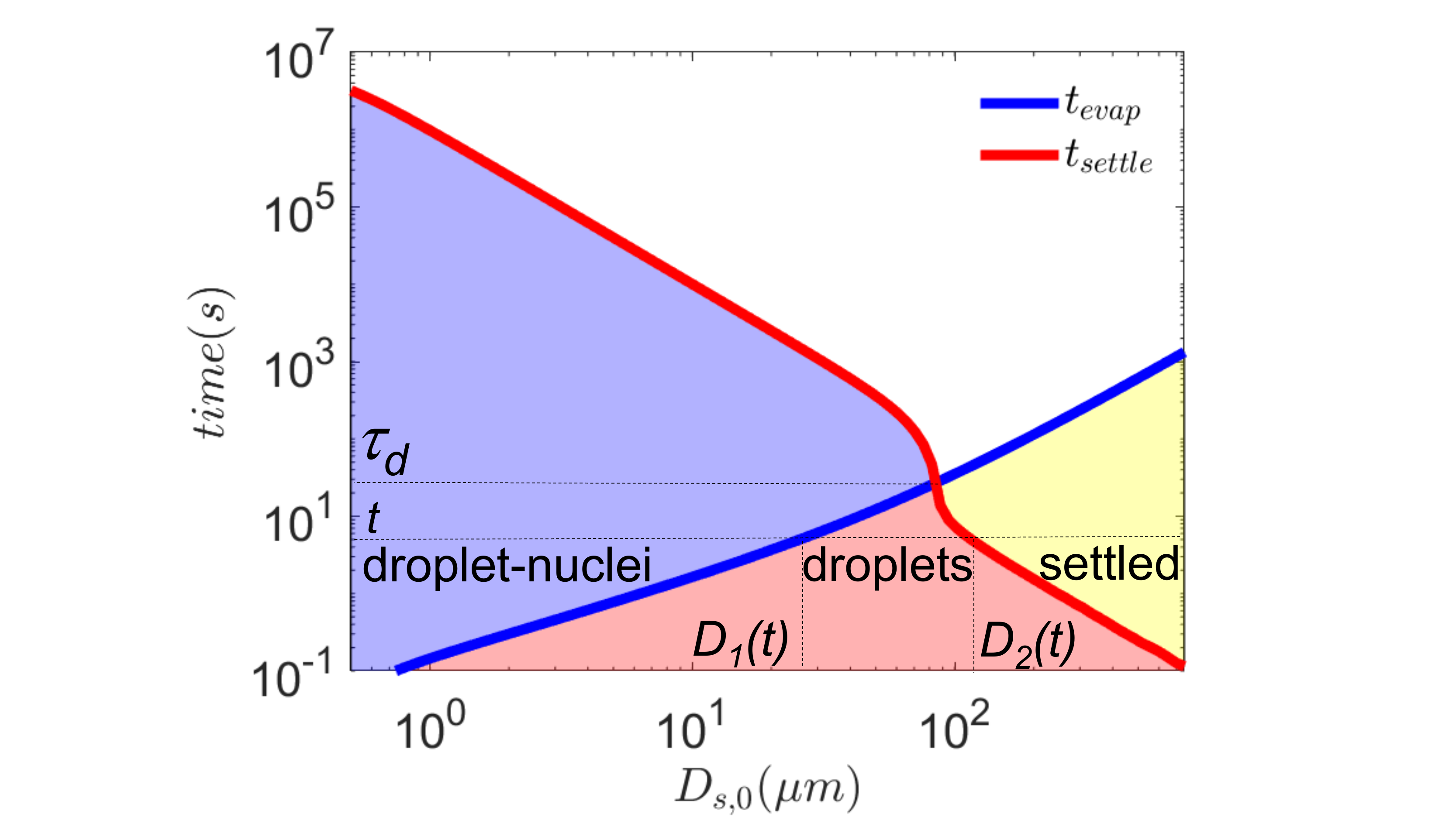}
\begin{centering}
\caption{Wells curves modified by accounting for droplet cloud aerodynamics and desiccation for 1\% NaCl-water droplets at $T_{\infty}=21.44^oC$ and $RH_{\infty}=50 \%$. After time $t$, the droplets with initial size $D_{s,0}<D_1(t)$ have been converted to dried droplet nuclei; those within $D_1(t) \leq D_{s,0} \leq D_2(t)$ are in liquid droplet state; $D_{s,0}>D_2(t)$ have settled and could be potential fomites. $\tau_d =22.87s$. The red, blue and the yellow shaded regions denote regimes of droplets, dried droplet nuclei and fomites, respectively}
\label{Fig:Mod_Wells}
\end{centering}
\end{figure}

For the SARS-CoV-2 virus, W{\"o}lfel reported \cite{wolfel2020virological} the average viral load in sputum to be $\rho_v = 7 \times 10^6$ copies/ml while the maximum is given by $\rho_{v,max}= 2.35 \times 10^9$ copies/ml.
Utilizing Eqn. \ref{Eq: Nvd}, we can define  $\mathcal{N}_{\alpha d}(t)$ - time dependent number of virions inhaled from droplets.
For an infection to occur, some non-zero number of active virion must be found in the droplets present in the total volume of air inhaled. Maximum time for $S$ to cross the volume with diameter $\sigma_{DS}$ is given by $t_{cross}=(\sigma_S + \sigma_D)/V_{DS}$ while number of breaths per unit time is $N_b \approx 16/60 s^{-1}$ \cite{barrett2019ganong} and volume inhaled per breath is $\mathcal{V}_b=(4/6)\pi\sigma_S^3$. Therefore total volume of air inhaled while crossing the respiratory cloud is $\mathcal{V}_a=(4/6)\pi N_b \sigma_S^3(\sigma_S + \sigma_D)/V_{DS}$. The fraction of virion population surviving within the droplets or dried-droplet nuclei at time $t$ is given by $\psi_{\beta}(t)$, and can be assumed to decay as $\psi_{\beta}(t) = (1/2)^{t/t_{\beta\frac{1}{2}}}$. $t_{\beta\frac{1}{2}}$ is the half-life of the SARS-CoV-2 virus in $d$ or $n$. While the half-life of the SARS-Cov-2 within aerosols, in general could be estimated from \cite{schuit2020airborne}, the distinction of the half-life of the virus within droplets or dried droplet nuclei is not yet available to our knowledge. While this information is indeed most critical, in view of its absence, for the present work we will mostly assume $\psi_d(t)=\psi_n(t)=\psi(t)$ and $t_{\beta\frac{1}{2}} = t_{\frac{1}{2}}$, unless specifically mentioned. At typical indoor conditions: $T_{\infty}=21.1^o C, RH_{\infty}=50 \%$ and with UV index $=1$ on a scale of 10, the $t_{\frac{1}{2}}=15.25$ minutes. Accounting for these, $\mathcal{N}_{\alpha d}(t)$ is given by Eqn. \ref{Eq:Nvd equation} below.

\begin{equation}
\mathcal{N}_{\alpha d}(t)= \frac{\pi \rho_v N_{t \alpha} N_b \sigma_S^3 (\sigma_S + \sigma_D(t)) \psi_d(t)}{12 V_{DS}(t)\sigma_D^3(t)} \int_{D_1(t)}^{D_2(t)} D^3f_{\alpha}(D)dD
\label{Eq:Nvd equation}
\end{equation}
Here, $D_1(t)$ and $D_2(t)$ are the minimum and maximum of initial droplet diameters, respectively, available in the droplet cloud after time $t$ as shown in Fig. \ref{Fig:Mod_Wells}. The non-linearity of the $t_{settle}$ vs. $D_{s,0}$ (in log-log plot of Fig. \ref{Fig:Mod_Wells}) occurs due to phase transition of the droplet population. Beyond $\tau_d$ all droplets have been converted into dried droplet nuclei. At time $t$ droplets with $D_{s,0}<D_1(t)$ have evaporated and those have been converted to dried droplet nuclei, while $D_{s,0}>D_2(t)$ have escaped by gravitational settling, and have been converted to potential fomites. Clearly, as $\sigma_D$ increases with time, $\mathcal{N}_{\alpha d}$ decreases due to dilution effects and also because droplet numbers are depleted by evaporation and settling. Since, the maximum evaporation time of the airborne droplets: $\tau_d<<t_{\beta\frac{1}{2}}$, the effect of virus half-life on $\mathcal{N}_{\alpha d}$ is negligible. Details of the methodology to derive the parameters concerning the respiratory jet and droplet dynamics from the conservation principles of mass, momentum, energy and species could be found briefly, in subsections \ref{Jet-puff-cloud} and \ref{Droplet dynamics}, respectively. Further details could be found in Chaudhuri et al. \cite{chaudhuri2020modeling}.

It is essential to note that in many diseases, uncertainty exists over transmission routes. For e.g. in case of Covid-19, the transmission by dried droplet nuclei, is not certain. As such, we do not know for sure if the SARS-CoV-2 virus survives within dried droplet nuclei and remain culturable \cite{WHOJune292020}, though there is evidence that some other virus do survive quite well inside dried droplet nuclei \cite{marr2019mechanistic}. Even if they do, their half-life and infection potential could be different w.r.t. to those inside droplets. Thus, as would be shown later it is essential to define different rate constants for the different transmission modes. Since the dried droplet nuclei is a product of the droplets, itself, the two routes are highly coupled and are not independent. 

For the droplet-nuclei, $\mathcal{N}_{\alpha n}$ is given by

\begin{equation}
\mathcal{N}_{\alpha n}(t)= \frac{\pi \rho_v N_{t \alpha} N_b \sigma_S^3 (\sigma_S + \sigma_D(t)) \psi_n(t)}{12 V_{DS}(t)\sigma_D^3(t)} \int_{0}^{D_n(t)} D^3f_{\alpha}(D)dD
\label{Eq:Pvn equation}
\end{equation}
$D_n(t)=D_1(t)$ if $t<=\tau_d$ and $D_n(t)=D_2(t)$ if $t>\tau_d$.
$\mathcal{N}_{\alpha n}(t)$ decreases with time due to increase in $\sigma_D$ with time, i.e. dilution effect as well due to virus half-life. 
$\mathcal{N}_{\alpha n}(t)$ decreases with time due to increase in $\sigma_D$ with time, i.e. dilution effect as well due to virus half-life. 

The generalized probability of infection $\mathcal{P}_{\alpha \beta}$ as a function of the infectious dose $\mathcal{N}_{\alpha\beta}$ can now be expressed as
\begin{equation}
\mathcal{P}_{\alpha \beta}(t)=1-e^{-r_v \mathcal{N}_{\alpha \beta}(t)}
\label{Eq:Pv alpha beta equation}
\end{equation}

The total probability of infection for the expiratory event $\alpha$ could be defined as 

\begin{equation}
\mathcal{P}_{\alpha}(t)=1-e^{-r_v \sum_{\beta}\mathcal{N}_{\alpha \beta}(t)}
\label{Eq:Pv sum equation}
\end{equation}

The form of Eqn. \ref{Eq:Pv sum equation} is based on the dose response model by Haas \cite{haas1983estimation}, which has been used by Nicas \cite{nicas1996analytical},  Sze To et al. \cite{sze2008methodology} and many other authors to calculate infection probability. Mathematically, it is also similar to the Wells-Riley equation \cite{riley1978airborne}, used by  Buonanno et al. \cite{buonanno2020estimation} to assess the aerosol risk of SARS-CoV-2 during talking and breathing. However, in contrast to these works, here, droplet cloud aerodynamics (sub-section \ref{Jet-puff-cloud}) coupled with detailed droplet evaporation-nuclei production mechanism (sub-section \ref{Droplet dynamics}) and droplet settling dynamics are utilized in a semi-analytical framework to calculate the time varying inhaled virion number and corresponding probability of infection, probably for the first time. Eventually, as shown later, this framework will be used to calculate basic reproduction number $\mathcal{R}_0$. As such the form of this equation is also validated by the results by Zwart et al. \cite{zwart2009experimental}, where $r_v$ is a constant for a particular virus. For this paper we will use $r_v=0.5$ such that inhaling at least 10 virions by $d$ and/or $n$ route would result in an infection probability $\mathcal{P}_{\alpha \beta} \approx 1$, unless specifically mentioned. In absence of this exact $r_v$ for the SARS-CoV-2 virus at the time of writing this paper, this is a guess. Hence, $r_v=0.05$ and $r_v=0.005$ corresponding to minimum infectious doses of 100 and 1000 virions, respectively will also be eventually explored near the end of the paper. 

\subsection{Aerodynamics of droplets and nuclei} \label{Jet-puff-cloud}
The droplets when ejected during respiratory events, follow the volume of exhaled air. Due to continuous entrainment the exhaled air volume grows in diameter, and as a result its kinetic energy decays with time. Bourouiba et al.  \cite{bourouiba2014violent} identified that for a short duration, the exhaled droplets evolve inside a turbulent jet, which transitions to a puff at later stage. Since the respiratory droplets or the dried nuclei experiences aerodynamic drag, it is essential to identify the evolution of the surrounding jet or puff. Based on literature \cite{abani2007unsteady, cushman2010turbulent, scorer1997dynamics} of transient turbulent jets and puff, the following evolution equations for the axial location, velocity and radial spread could be used:
\begin{equation}
\begin{aligned}
  x_{j}(t) & =\left(\frac{12}{K}\right)^{1/2}\left(U_{j,0} R_{j,0}\right)^{1/2}t^{1/2},\\
  U_{j}(t) & =\frac{6U_{j,0} R_{j,0}}{Kx_j(t)},\\
  R_{j}(t) & =R_{j,0} + (x_j(t)-x_{j,0})/5 
\end{aligned}
\label{eq:jet_eqn}
\end{equation}
and
\begin{equation}
\begin{aligned}
  x_{pf}(t) & =\left(\frac{3m}{a}\right)R_{pf}(t),\\
  U_{pf}(t) & =U_{pf,0}\left(\frac{3mR_{pf,0}}{4aU_{pf,0}t}\right)^{3/4}, \\
R_{pf}(t) & =R_{pf,0}\left(\frac{4aU_{pf,0}t}{3mR_{pf,0}}\right)^{1/4} 
\end{aligned}
\label{eq:puff_eqn}
\end{equation}
where subscript $j$ and $pf$ denote jet and puff, respectively. $R_0$ and $U_0$ are the radius and axial velocity at a distance $x_0$. $K$ is a characteristic constant for turbulent jet and is reported to be 0.457 \cite{abani2007unsteady}. At the inception of the respiratory event ($t=0$), the jet is assumed to have a velocity $U_{j,0}=10m/s$ and a radius $R_{j,0}=14mm$ - the average radius of human mouth. For analytical tractability, we assume that all droplets of all sizes are ejected at time $t=0$ and would not consider time variation in ejection of the droplets. This is a safe assumption since the expiratory event like cough, lasts less than a second and the turbulence of the jet and the air entrained will rapidly disperse the ejected droplets into the jet/puff in any case. The characteristic constants for puff are $a\approx2.25$ and $m=(x_{p,0}a)/(3R_{p,0})$ \cite{cushman2010turbulent}. Since the continuous ejection of air from mouth lasts only for the duration of the corresponding respiratory event, the jet behavior persists only for this period. Beyond this time ($\approx 1s$ \cite{han2013characterizations}), puff behavior is observed. Hence the velocity and the radial spread of the air surrounding the exhaled droplets will be
\begin{equation}
\begin{aligned}
  U_g =
    \begin{cases}
      U_j(t) & t\le 1s\\
      U_{pf}(t) & t>1s
    \end{cases}  \\ 
     R_g =
    \begin{cases}
      R_j(t) & t\le 1s\\
      R_{pf}(t) & t>1s
    \end{cases}   
     \label{eq:gas_vel}
     \end{aligned}
\end{equation}
The horizontal displacement ($X_{p}$) of the exhaled droplet and its instantaneous velocity ($U_p$) due to the drag can be solved with \cite{Sirignano_2010}:
\begin{equation}
\begin{gathered}
    dX_p/dt=U_p\\
    dU_p/dt=\left(\frac{3 C_D \rho_v}{8R_s \rho_l}\right) |U_g - U_p| (U_g - U_p)
    \end{gathered}
     \label{eq:drag}
\end{equation}
Here, $\rho_v$ and $\rho_l$ are the vapor and liquid phase densities, respectively; $R_s$ is the instantaneous radius of the droplet and $C_D$ is the drag coefficient. We can assume $C_D=24/Re_p$ for the gas phase Reynolds number, $Re_p=(2\rho_v|U_g - U_p|R_s)/\mu_g<30$ \cite{Sirignano_2010}. $Re_p $ for the respiratory droplets are typically less than $0.1$. 
At the time of ejection ($t=0$) from respiratory cavities, the droplets are assumed to have a velocity ($U_{p,0}$) close to that of the surrounding air ($U_{j,0}$), and hence $Re_p \approx 0$. Thus, for $t=0$, we use $U_{j,0}-U_{p,0}=0.01U_{j,0}$. 
$t_{settle}(D_{s,0})$ - the time for a droplet with initial diameter $D_{s,0}$ to fall a height of $1.8m$ is calculated using Stokes' settling velocity.  Mathematically $t_{settle}$ is obtained by Eqn. \ref{Eqn:Settling time} 
\begin{equation} \label{Eqn:Settling time}
(18 \mu)^{-1}\int_{0}^{t_{settle}} (\rho_l(t) - \rho_v)g D_s^2(t) dt = h_0	
\end{equation}
By solving Eqns. \ref{eq:jet_eqn}-\ref{eq:drag} over the droplet and nuclei lifetime, the axial distance traveled by them, $X_D$, which is the distance of the center of the cloud, can be evaluated. As the velocity of the individual droplets approach the surrounding gas velocity within a very short time, we assume the absolute instantaneous velocity of the droplet/nuclei cloud is given by $V_D=U_g$ from Eqn. \ref{eq:gas_vel}. Since the droplets and nuclei are dispersed within the jet/puff, the diameter of the droplet cloud ejected by $I$ can be approximated as twice the radial spread of the exhaled air, $\sigma_D(t) = 2R_g(t)$. 

The exhaled volume of air is initially at a temperature ($T_{g,0}=33.25^oC$) and vapor mass fraction ($Y_{1,g,0}$ corresponding to $RH_0=71.6 \%$) different from the ambient. The values mentioned are averaged quantities measured over several subjects according to Mansour et al. \cite{mansour_2020}. The instantaneous temperature and vapor mole fraction that the droplet would encounter as its own ambient are the temperature ($T_{g}(t)$) and vapor mass fraction ($Y_{1,g}(t)$) of this volume of air during its evolution. This can be expressed with the following scaling relation \cite{abramovich_2003}
\begin{equation}
\begin{gathered}
    \frac{\Delta T_g}{\Delta T_{g,0}} =\frac{\Delta Y_{1,g}}{\Delta Y_{1,g,0}} = \frac{U_g}{U_{g,0}},
    \end{gathered}
     \label{eq:jet-T-Y}
\end{equation}
where $\Delta T_g=T_g-T_\infty$ and $\Delta Y_{1,g}=Y_{1,g}-Y_{1,\infty}$. 

\subsection{Droplet evaporation} \label{Droplet dynamics}
\begin{figure}[ht]
\includegraphics[trim=1cm 8.5cm 1cm 7.5cm,clip,width=0.8\textwidth]{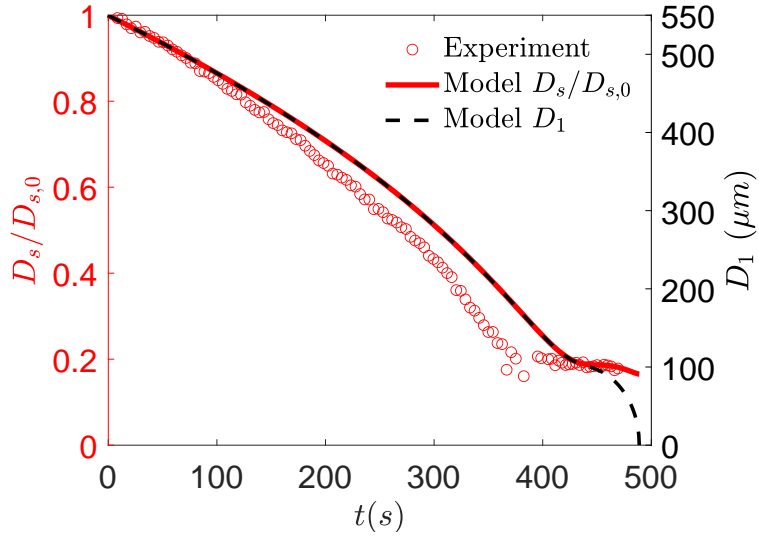}
\begin{centering}
\caption{Comparison of the model output for 1\% NaCl-water droplets with human saliva droplet experimental data averaged, over three runs. $D_s$ (normalized with $D_{s,0}$ shown in left ordinate) is the effective diameter of the droplet accounting for both the solute and solvent while $D_1$ (shown in right ordinate) is the effective diameter only accounting for the solvent (water) mass of the droplet.}
\label{Fig:Validation}
\end{centering}
\end{figure}

In this paper we use $1\%$ NaCl-water droplets as the model respiratory droplet and adopt a slightly revised evaporation model (with respect to that presented in \cite{chaudhuri2020modeling}) for predicting the droplet evaporation time $t_{evap}$. It is to be recognized that the droplets are surrounded by exhaled air volume as described in sub-section \ref{Jet-puff-cloud} and hence, it serves as the ``ambient condition'' for the droplet. The evaporation mass flux for quasi-steady state condition is given by
\begin{equation}
\begin{split}
    \dot{m}_1=-4\pi\rho_v D_v R_s ln(1+B_M) \\
    \dot{m}_1=-4\pi\rho_v \alpha_g R_s ln(1+B_T)
\end{split}
    \label{eq:mdot}
    \end{equation}
Here, $\dot{m}_1$ is droplet mass loss rate due to evaporation, $R_s$ the instantaneous droplet radius, $\rho_v$ is density of water vapor and $D_v$ is the binary diffusivity of water vapor in air, $\alpha_g$ is the thermal diffusivity of surrounding air.    
$B_{M}=(Y_{1,s}-Y_{1,g})/(1-Y_{1,s})$ and $B_{T}=C_{p,l}(T_{s}-T_{g})/h_{fg}$ are the Spalding mass transfer and heat transfer numbers, respectively. $Y$ is mass fraction with numerical subscripts 1, 2 and 3 denoting water, air and salt respectively. Additionally, subscript $s, g, \infty$ denote location at droplet surface, surrounding gas and at very far field ambient, respectively.  $h_{fg}$ and $C_{p,l}$ are the specific latent heat of vaporization and specific heat of the droplet liquid, respectively. Unlike in a pure water droplet, vapor pressure at the surface of droplets with non-volatile dissolved substances as in respiratory droplet/salt solution droplets could be significantly suppressed. Raoult's Law provides the modified vapor pressure at the droplet surface for binary solution, $P_{vap}(T_s,\chi_{1,s})=\chi_{1,s}P_{sat}(T_s)$, where $\chi_{1,s}$ is the mole fraction of evaporating solvent (here water) at droplet surface in the liquid phase \cite{Sirignano_2010} and $\chi_{1,s} = 1-\chi_{3,s}$. The far field vapor concentration, on the other hand, is related to the relative humidity of the ambient.
Considering the effects of Raoult's law and relative humidity, the vapor concentrations at droplet surface and at far field can be expressed as: 
\begin{equation} \label{eq:Yfs}
\begin{split}
Y_{1,s}=\frac{P_{vap}(T_s, \chi_{1,s})M_1}{P_{vap}(T_s,\chi_{1,s})M_1 + (1-P_{vap}(T_s,\chi_{1,s}))M_2}  
\end{split}
\end{equation}
where $M_1$, $M_2$ are molecular weights of water and air, respectively. Instantaneous $Y_{1,g}$ is evaluated from Eq. \ref{eq:jet-T-Y}. The latent heat required for evaporation, is provided by the droplet's internal energy and/or surrounding ambient. It has been verified that the thermal gradient in the liquid phase is rather small. Therefore, neglecting the internal thermal gradients $T_s$ is obtained from the energy balance
\begin{equation} 
    mC_{p,l}\frac{\partial T_s}{\partial t}=-k_gA_s\frac{\partial T_s}{\partial r}|_s + \dot{m}_1 h_{fg}
    \label{eq:HT}
\end{equation}
where, $T_s$ is instantaneous droplet temperature; $m=(4/3)\pi\rho_l R_s^3$ and $A_s=4\pi R_s^2$ are the instantaneous mass and surface area of the droplet; $\rho_{l}$ is the density of the binary mixture of salt (if present) and water and $k_g$ is the conductivity of air surrounding the droplet. $\frac{\partial T}{\partial r}|_s$, is the thermal gradient at the droplet surface and can be approximated as $(T_s-T_g)/R_s$. Due to continuous loss of water, the solution would become supersaturated in most occasions leading to the onset of crystallization. The crystallization kinetics is modeled with a one-step reaction \cite{naillon2015evaporation,derluyn2012salt}. The validation of the model (1\% NaCl-water solution) with saliva droplet experiments (average of three runs) from a healthy subject is shown in Fig. \ref{Fig:Validation}. The experiments were performed in a contact free condition in  an acoustic levitator at $T_{\infty}=28^o C$ and $RH_{\infty}=41\%$. The reader is referred to ref. \cite{chaudhuri2020modeling} for details of the experimental configuration. Here, a difference of upto $15\%$ could be found for the different stages of droplet drying, between the model prediction and the saliva droplet drying curve from experiments. It should be noted that human saliva contains mucus, varieties of salts and electrolytes along with compositional variations, which are difficult to model very accurately.

\subsection{\textit{Ab initio} rate constants for SEIR model}
With the probability of different transmission routes identified, we proceed to evaluate the respective ``rate constants''. A theoretical framework that explicitly connects respiratory droplets to the pandemic dynamics is rarely available. Within the framework of the well known SIR-model, the model constants proposed by Stilianakis and Drossinos \cite{stilianakis2010dynamics} included overall droplet cloud features like number of droplets per unit volume of the cloud, but did not include crucial physics like cloud aerodynamics, evaporation or crystallization dynamics that lead to droplet-nuclei formation. As such these control the time evolution of the droplet cloud and as shown later the spatio-temporal evolution of the cloud and the constituent droplets play a major role in determining the critical rate constants of the problem. A model connecting the macro-scale pandemic dynamics with the micro-scale droplet physics accounting for droplet-cloud aerodynamics, evaporation and crystallization physics has been recently presented by Chaudhuri et al. \cite{chaudhuri2020modeling}. Drawing inspiration from the well known molecular collision theory of reactions due to collisions, a chemical reaction mechanism was obtained where three elementary reactions described the pandemic evolution. Adopting the notations of the SEIR model, one of the reaction rate constants that determined conversion of a susceptible individual $S$ to an exposed individual $E$, upon contact with the droplet cloud $D$ ejected by the infectious person $I$, was denoted by $k_1$ (or $k_{1,o}$ as opposed to the new rate constant to be defined here) and was called the infection rate constant. This $k_{1,o}$ was modeled using the molecular collision theory \cite{law_2006}. From Chaudhuri et al. \cite{chaudhuri2020modeling}, the expected number of collisions per unit time between $S$ and $D$ of $I$ is given by $\pi \sigma_{DS} ^2 V_{DS} n_I n_S$ resulting in the infection reaction $I + S \rightarrow{} I + E$. $n_I$ and $n_S$ are the number of infected $I$ and susceptible $S$ people in unit volume. The infection rate constant of this reaction is then given by 
\begin{equation}
k_{1,o} = \pi n_{total} \sigma_{DS} ^2 V_{DS} (\tau_d / t_c)
\label{Eq:k_1 equation}
\end{equation}
$\sigma_{DS}$ is the jet/puff diameter, which is also assumed to be the diameter of the droplet cloud. $\vec{V}_{DS}$ is the relative velocity of the droplet cloud $D$ w.r.t. $S$ while $\tau_d$ is the droplet lifetime. $V_S$ can be approximated as the preferred walking speed which according to \cite{karamouzas2014universal, weidmann1993transporttechnik} equals 1.3$\pm 0.3$m/s, and hence for the current study we will assume $V_S=1.3$m/s. $t_c$ is the average time period between two expiratory events. $t_c$ is calculated as $t_c=3600 \times 24/N_{\exp}$ where $N_{\exp}$ is the average number of infecting expiratory events per person per day. We assumed $N_{exp}=3$, from the measured coughing frequency of 0-16 in normal subjects \cite{hsu1994coughing}. 

The above expression Eqn. \ref{Eq:k_1 equation} is limited by several simplifying assumptions. In this paper, we develop a comprehensive model beyond these limitations, which is also rendered capable of delineating the relative dominance of the different disease transmission pathways. In particular, in the following we derive a new rate constant accounting for i) the finite viral loads, finite viral lifetime and the corresponding probability of infection ii) the evolution of the collision volume with time iii) transmission by droplets of any sizes and the corresponding dried droplet nuclei and iv) inhomogeneity of infection spreading.  Furthermore, in this paper we generalize the infection rate constant equation to account for transmission by any expiratory event. To that end, a generalized reaction mechanism that accounts for different modes of infection transmission as well as different form of expiratory events are presented. This is followed by a comprehensive modeling of the individual infection rate constants, following which we arrive at an overall infection rate constant. 
In view of the above discussion, the basic reaction mechanism of \cite{chaudhuri2020modeling} could be generalized to a comprehensive one where almost all possible expiratory events and modes of transmission could be included to yield: 

$$ \ \ \ce{ \textit{S} + \textit{I} ->[$k_{1,\alpha\beta}$] \textit{E} + \textit{I} } \ \ \ \ \ \   [R1_{\alpha\beta}] $$
$$\ce{ \textit{E} ->[$k_2$] \textit{I}  } \ \ \ \ \ \ \ \ \ \ \ \ \ \ \ \ \ \ \ \ [R2] $$
$$\ce{ \textit{I} ->[$k_3$] 0.97\textit{R} + 0.03\textit{X}  } \ \ \ \ [R3] $$

In $[R1_{\alpha\beta}]$, $\alpha$ varies over different expiratory events, namely breath, cough, sing, sneeze, talk, while 
$\beta$ varies over different modes of transmission, namely droplet, droplet nucleus, fomite. Thus $[R1_{\alpha\beta}]$ essentially represents 15 reactions. The rate constants of these individual reactions are defined in Table \ref{tab:1}.

\begin{table}[h!]
\begin{center}
\begin{tabular}{ | m{2cm} | m{2cm}| m{2cm} | m{2cm} | } 
\hline
$k_{1,\alpha\beta}$ & droplet & nucleus & fomite \\ 
\hline
breath  & $k_{1,bd}$ & $k_{1,bn}$ & $k_{1,bf}$ \\ 
\hline
cough  &  $k_{1,cd}$ &  $k_{1,cn}$ &  $k_{1,cf}$ \\ 
\hline
sing  & $k_{1,gd}$ & $k_{1,gn}$ & $k_{1,gf}$\\ 
\hline
sneeze  & $k_{1,sd}$ & $k_{1,sn}$ & $k_{1,sf}$\\ 
\hline
talk  & $k_{1,td}$ & $k_{1,tn}$ & $k_{1,tf}$ \\ 
\hline
\end{tabular}
\caption{\label{tab:1}Infection rate constants for different expiratory events and modes of transmission}
\end{center}
\end{table}


Here, each of the parameters should be obtained for the respective combination of $\alpha,\beta$. Including $[R2]$ and $[R3]$, in total there are $15+2=17$ reactions to be included in a complete model. As such further granularity could be added by adding a location parameter $\gamma$. In that case we can have $k_{1,\alpha\beta\gamma}$, where $\gamma$ could represent home, school, office, transport etc. In this paper, we will only consider two selected transmission modes: cough-droplets and cough-dried droplet nuclei with the rate constants $k_{1,cd}$ and $k_{1,cn}$ as shown in Table \ref{tab:1}. These are expected to play the more dominant roles in disease transmission. However, the approach here could be used for any other transmission routes too, with the corresponding droplet size distribution. For Covid-19 fomites are being considered a secondary source of infection and needs to be dealt separately.

Individual rate constants will allow us to delineate the different modes of transmission on average. Furthermore, definition of individual rate constants enables quantitative investigation of the relative dominance of each mode of transmission. This constitutes one of the major goals of the paper. The infection rate constants are generalized by inclusion of the probability for infection $\mathcal{P}_{\alpha \beta}$, averaging the collision volume over a characteristic time alongside including the dried droplet nuclei mode of transmission, in addition to the droplet mode of transmission. The revised rate constant for any expiratory event $\alpha$, vector of transmission $\beta$ and location $\gamma$ is given by Eqn. \ref{Eq:k1_revised}
\begin{equation}
k_{1,\alpha \beta \gamma} = \frac {\pi n_{total,\gamma}}{t_c} \int_{0}^{\tau} \sigma_{DS} ^2(t) V_{DS}(t) \mathcal{P}_{\alpha \beta}(t) dt
\label{Eq:k1_revised}
\end{equation}

Specifically, by utilizing Eqn. \ref{Eq:Nvd equation} and Eqn. \ref{Eq:Pv alpha beta equation}, the rate constant for the droplet mode of transmission $d$, ejected during any expiratory event $\alpha$ could be calculated as

\begin{equation}
k_{1,\alpha d \gamma} = \frac{\pi n_{total, \gamma}}{{t_c}}  \int_{0}^{\tau_d} \sigma_{DS}^2(t) V_{DS}(t) \mathcal{P}_{\alpha d}(t) dt
\label{Eq:k1_alpha_d equation}
\end{equation}

For the droplet-nuclei, Eqn. \ref{Eq:Pvn equation} and and Eqn. \ref{Eq:Pv alpha beta equation} yields  the rate constant $k_{1,\alpha n \gamma}$

\begin{equation}
k_{1,\alpha n \gamma} = \frac{\pi n_{total, \gamma}}{{t_c}}  \int_{0}^{\infty} \sigma_{DS}^2(t) V_{DS}(t) \mathcal{P}_{\alpha n}(t) dt
\label{Eq:k1_alpha_n equation}
\end{equation}

Note, that we introduced a new parameter $\psi(t)$ to calculate $\mathcal{P}_{\alpha \beta}$ which denotes the fraction of the infectious virion population active within the dried droplet nuclei population at time $t$. As $t \xrightarrow{} \infty$, $\psi(t) \xrightarrow{}0$. Thus the integration is performed by upto about $max(t_{evap}) \sim \mathcal{O}(1000s)$ - the largest evaporation time of the droplet set considered. Details on the survivability of specific SARS-CoV-2 inside dried droplet nuclei is not known. Hence, for now we will assume $\psi(t)$ is independent of $d$ and $n$, except when we will estimate its sensitivity in specific cases.
According to the reaction mechanism given by $[R1_{\alpha\beta}], [R2], [R3]$, $E$ is formed by several parallel pathways. Therefore, the corresponding rate constants become additive. Hence, the location ($\gamma$) dependent infection rate constant can be defined as:

\begin{equation}
k_{1,\gamma} = \sum_{\alpha,\beta} k_{1,\alpha \beta \gamma}
\label{Eq:k1 overall}
\end{equation}

While the rate constant $k_{1,\gamma}$ is derived from first principles, it still results in same infection rate constant for a given set of ambient temperature $T_{\infty}$, $RH_{\infty}$ and population density. As shown by Lloyd-Smith et al. \cite{lloyd2005superspreading} the individual infectiousness distribution around the basic reproduction number is highly skewed. This suggests that a small fraction of infected individuals ``superspreaders'' are responsible for a large number of infections. Hence, the final challenge of this modeling effort is to include this effect. Such ``superspreading'' events could be results of i) high local population density  $n_{total,\gamma}$, ii) highly mobile infected individuals and iii) most importantly, high viral loading of the ejected respiratory droplets $\rho_v$. We will see that large viral loading $\rho_v=\rho_{v,max}$ leads to very high infection probability which would lead to large $k_{1,\gamma}$. Since the rate constant is directly proportional to $n_{total,\gamma}$, its effect is understandable. Thus, effect resulting from mobility needs to be accounted. 

Understanding and modeling human mobility at both individual level as well as at a population level has garnered recent interest. See a recent review by Barbosa et al.\cite{barbosa2018human} for a detailed exposition on this topic. Kolbl and Helbing \cite{kolbl2003energy} used statistical data of the UK National Travel Surveys collected for 26 years by the Social Survey Division of the Office of Population Census and Surveys to arrive at a generalized distribution of human daily travel behavior. They showed that for different modes of transport $i$ ranging from walking, cycling, car driving etc., the travel time $t_t$ normalized by the average travel time for the corresponding mode of travel $\bar{t}_{t,i}$ and defined as $\tau_{t,i}=t_t/\bar{t}_{t,i}$, a common distribution for $\tau_{t,i}$, irrespective of the mode of transport could be obtained. This was also argued from an energy point of view, where $E_i/\bar{E}=\tau_{t,i}$; $\bar{E}=615$ kJ per person per day - the average travel energy budget of the human body according to \cite{kolbl2003energy}. In any case the pdf, $g_{\tau_t}$, after dropping the $i$ given its universality is

\begin{equation}
g(\tau_{t})=N'exp(-\alpha/\tau_{t} - \tau_{t}/\beta).
\label{Eq:gtau distributio}
\end{equation}
The following constants were provided, $N=N'/\bar{E}=2.5, \alpha=0.2, \beta=0.7$ for the universal curve \cite{kolbl2003energy}. 

Given two infected people $I$, it is reasonable to expect that the one with the higher mobility has more chance to infect others since they have greater exposure to the population and can infect people at different locations, all other conditions remaining fixed. Therefore, we can assume that the final infection rate constant should be proportional to $\tau_{t}$.

The corresponding infection rate constant summed over all possible types of expiratory event $\alpha$, transmission mode $\beta$ and location $\gamma$ is thus given by 
\begin{equation}
k_{1,\tau_{t}} = \tau_{t} \sum_{\alpha,\beta,\gamma} k_{1,\alpha \beta \gamma}.
\label{Eq:k1t overall}
\end{equation}
Clearly $k_1$ is now a function of two random variables $\tau_t$,  $n_{total,\gamma}$ and the extreme individual realization of each could correspond to the superspreading events.

Finally, the average, overall infection rate constant $k_1$ is given by
\begin{equation}
k_{1} = \int_{0}^{\infty} \tau_{t} g(\tau_{t}) \sum_{\alpha,\beta,\gamma} k_{1,\alpha \beta \gamma} d\tau_t = \sum_{\alpha,\beta,\gamma} k_{1,\alpha \beta \gamma}.
\label{Eq:k1t overall}
\end{equation}
This is because $\int_{0}^{\infty} \tau_{t} g(\tau_{t}) d\tau_t=1$ and that $\sum_{\alpha,\beta,\gamma} k_{1,\alpha \beta \gamma}$ is independent of $\tau_t$. In the rest of the paper we will mostly focus on this ensemble averaged rate constant $k_1$. With the framework established, the individual effects of mobility and population inhomogeneity could be taken up in future works. 

From the reactions $[R1_{\alpha\beta}], [R2], [R3]$ we can obtain the set of ordinary differential equations of the SEIR model, that would govern the evolution of $[I], [E], [S], [R]$ and $[X]$, where the rates constants appear as respective coefficients. Square brackets denote number of the particular population type normalized by the total population, for e.g. $[I]=n_I/n_{total}$.
\begin{equation} 
\begin{aligned} 
\frac{d[I]}{dt} = k_2[E] - k_3[I]  \\
\frac{d[E]}{dt} = k_{1}[I][S] - k_2[E]
\\ 
\frac{d[R]}{dt} = 0.97k_3[I]
\\
\frac{d[X]}{dt} = 0.03k_3[I]
\\
[S]+[E]+[I]+[R]+[X]=1
\end{aligned}
\label{Eq:SEIR_ODE}
\end{equation}


\section{Results and Discussion} 

\begin{figure}[ht!]
\begin{centering}
\begin{subfigure}{0.47\textwidth}
\includegraphics[trim=3cm 8.5cm 3cm 7.5cm,clip,width=\textwidth]{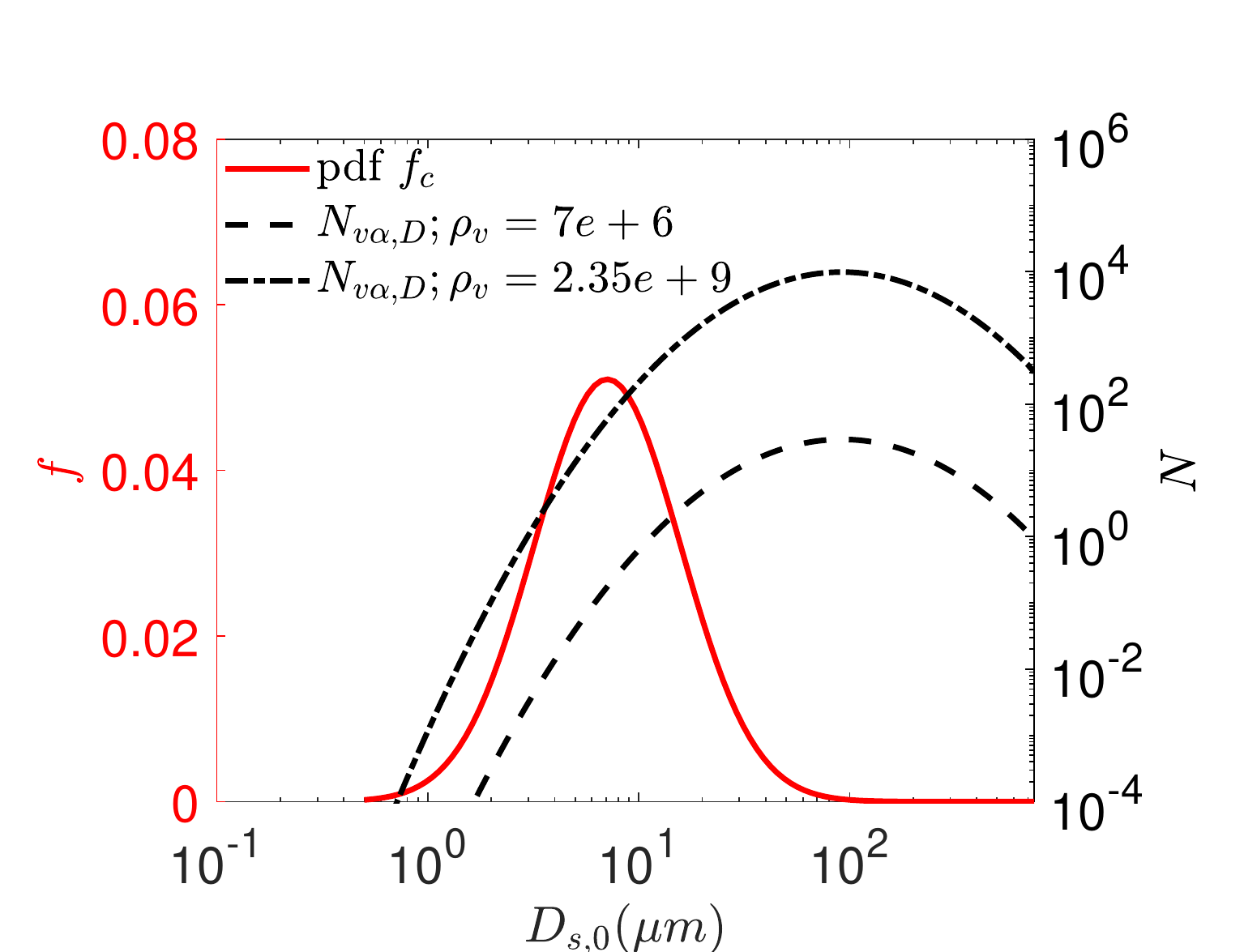}
\caption{}
\label{Fig:Dropletsize(a)}
\end{subfigure}
\begin{subfigure}{0.47\textwidth}
\includegraphics[trim=3cm 8.5cm 3cm 7.5cm,clip,width=\textwidth]{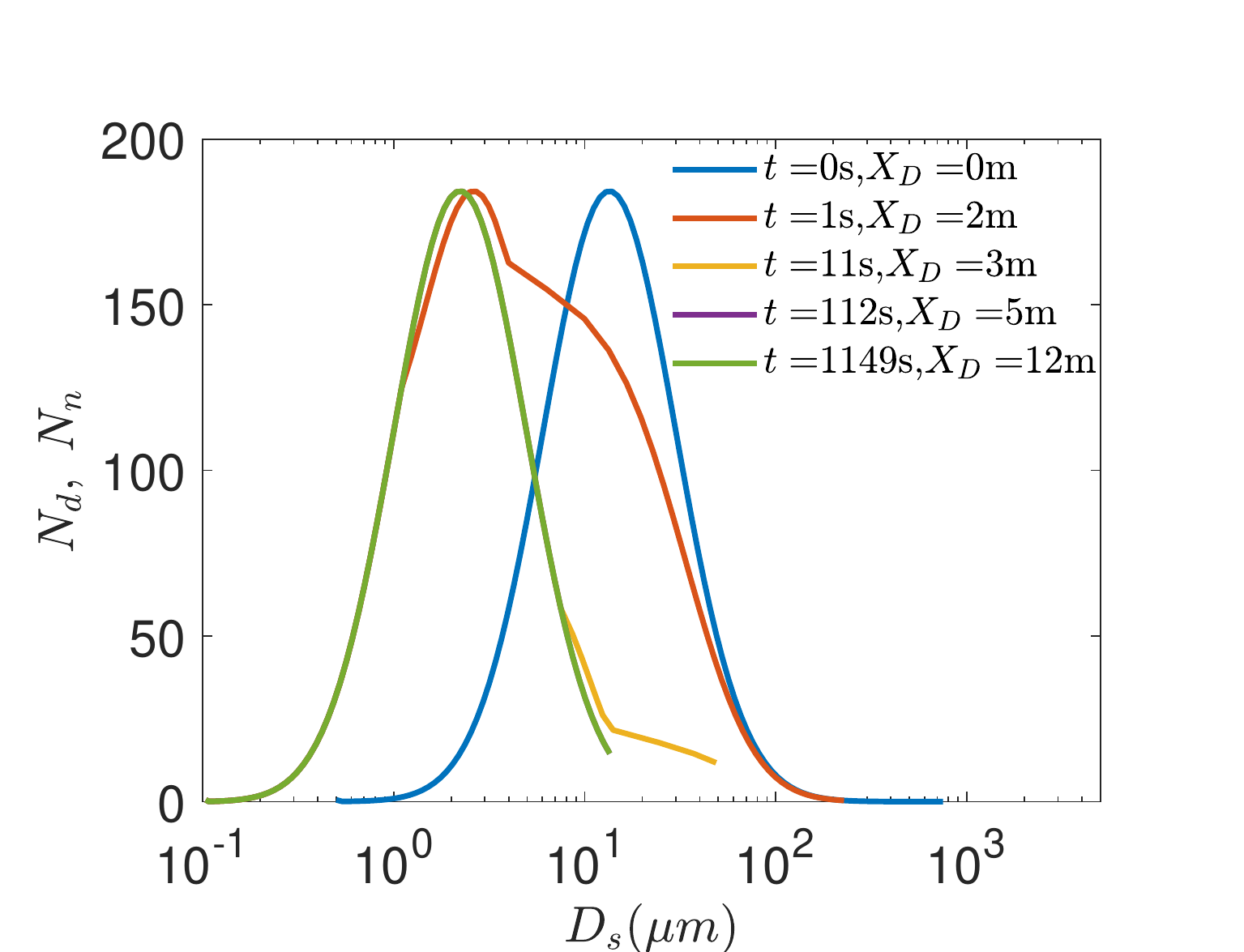}
\caption{}
\label{Fig:Dropletsize(b)}
\end{subfigure}
\caption{(a) Probability Density Function (PDF) $f$ of droplet diameter for cough \cite{duguid1946size} and number of virions $N$ as a function of the initial droplet size $D_{s,0}$ at $t=0$. Average and maximum viral load $\rho_v = 7 \times 10^6$ and $\rho_v = 2.35 \times 10^9$ copies/ml of SARS-CoV-2, respectively are assumed from \cite{wolfel2020virological}. (b) Droplet and/or dried droplet nuclei size distributions at different time $t$ and distance $X_D$ from the origin of the respiratory jet, for $T_{\infty}=21.44^o C, \ RH_{\infty}=50 \%$. The curves at the last three time instants end abruptly due to loss of droplets due to settling beyond that particular diameter. For $t>\tau_d =22.87s$ or $X_D > 3.52m$ all airborne droplets have been desiccated to the corresponding droplet nuclei. Size distribution remains invariant for $t>\tau_d$.}
\label{Fig:Dropletsize}
\end{centering}
\end{figure}

From the measurements by Duguid \cite{duguid1946size}, the droplet size distribution from cough could be described using a lognormal distribution. The initial distribution $f_c$ and the number of virions present in each droplet size for the average viral load $\rho_v=7 \times 10^6$ copies/ml is shown in Fig. \ref{Fig:Dropletsize(a)}. Total number of droplets ejected  = 5000 \cite{duguid1946size}. Figure 
\ref{Fig:Dropletsize(b)} shows the time evolution of the droplet number distribution as a function of the instantaneous diameter $D_s$. The shift of the distribution to the left i.e. towards smaller $D_s$ is an effect of evaporation. Also, the right branch of the number distribution gets eroded due to settling of the larger droplets. At the conditions of interest: $T_{\infty}=21.44^oC$ and $RH_{\infty}=50 \%$ the modal diameter of the droplet nuclei is $2.3 \mu m$ at $t=1050$s starting from an initial modal diameter of $13.9 \mu m$ at $t=0$s. Note that here, by modal diameter, the $D_s$ corresponding to the peak of the histogram shown in Fig. \ref{Fig:Dropletsize(b)} is referred. Interestingly, since small droplets evaporate fast, the left branch (small sizes) of the distribution shifts fast to further smaller sizes. A droplet with initial diameter $D_{s,0}=10 \mu m$ is reduced to $D_s=1.96 \mu m$ within $t=0.42s$. As such, for the entire droplet set, a modal diameter of $2.7 \mu m$ which is within $22 \%$ of the final modal diameter, is achieved within $t=1s$ or within a distance of $X_D=1.8m$ from the origin of the respiratory jet. $X_D$ denotes the distance of the center of the respiratory jet/puff (with a diameter of $\sigma_D$) from its origin. Within $t=10.6s$, $X_D=2.88m$, the droplet size distribution is very close to the final distribution. Due to this sharp reduction in droplet size due to evaporation (for $RH_{\infty} < 85 \%$) combined with settling of large droplets, practically, for most of the time, the disease appears to be transmitted by droplets/nuclei of instantaneous diameter less than $10 \mu m$, the most probable instantaneous diameter being between $2.14 \mu m$ and $2.7 \mu m$. However, it is to be noted that this diameter could be 5-6 times smaller than the initial ejected diameter of the droplet $D_{s,0}$. In a viewpoint article Fennelly \cite{fennelly2020} reported that for most respiratory infections, the smaller droplets ($<5 \mu m$ and collected at a finite distance from the origin of the respiratory spray) were found to be pathogenic. Furthermore Chia et al. \cite{chia2020detection} reported PCR positive SARS-CoV-2 particles with sizes $>4\mu m$ and also between $1-4\mu m$ from air samples collected. Thus, our results appear to be consistent with these clinical research findings. 

\begin{figure}[ht!]
\begin{centering}
\begin{subfigure}{0.470\textwidth}
\includegraphics[trim=3cm 8.5cm 3cm 7.5cm,clip,width=\textwidth]{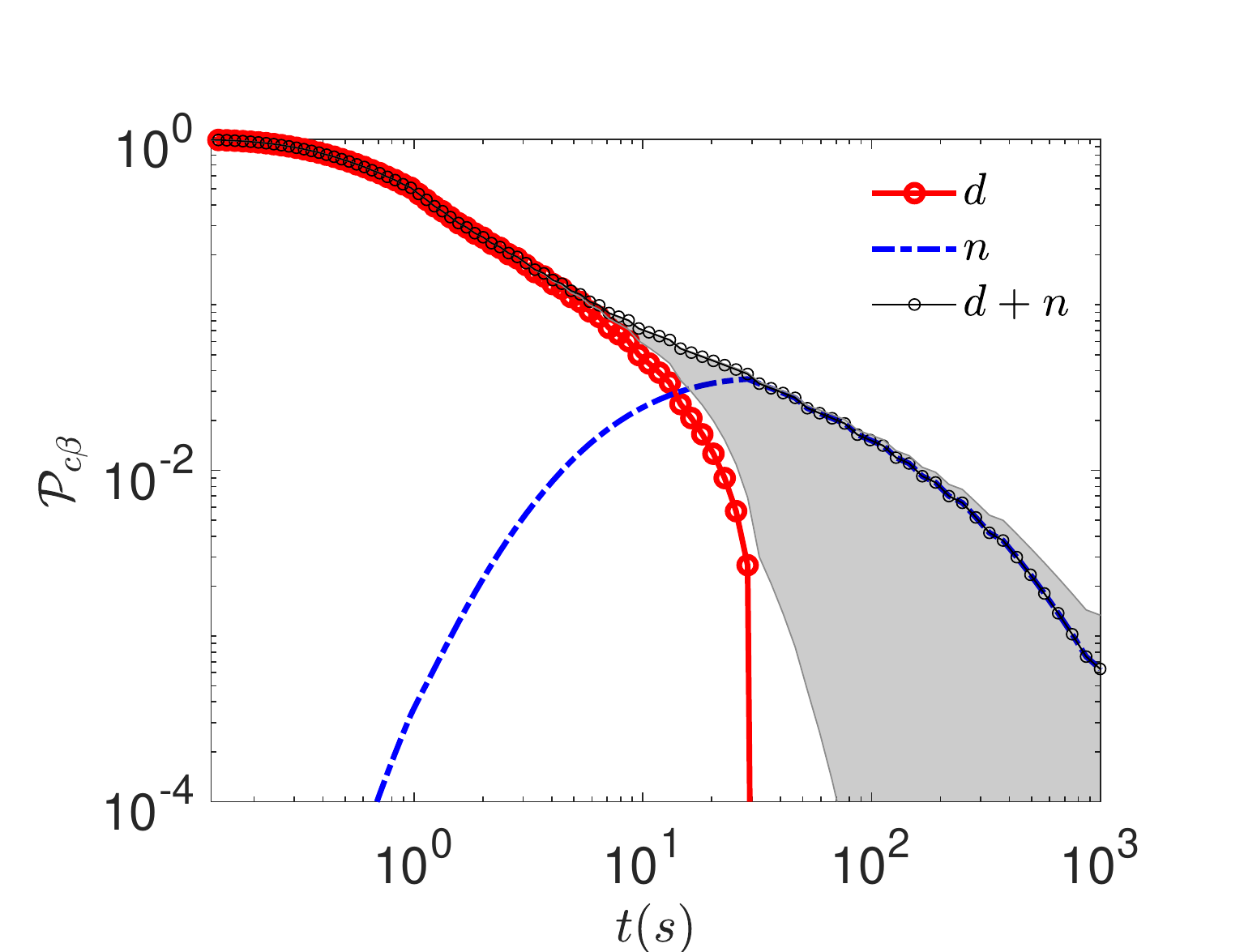}
\caption{}
\label{Fig:Pvd_Pvn(a)}
\end{subfigure}
\begin{subfigure}{0.470\textwidth}
\includegraphics[trim=3cm 8.5cm 3cm 7.5cm,clip,width=\textwidth]{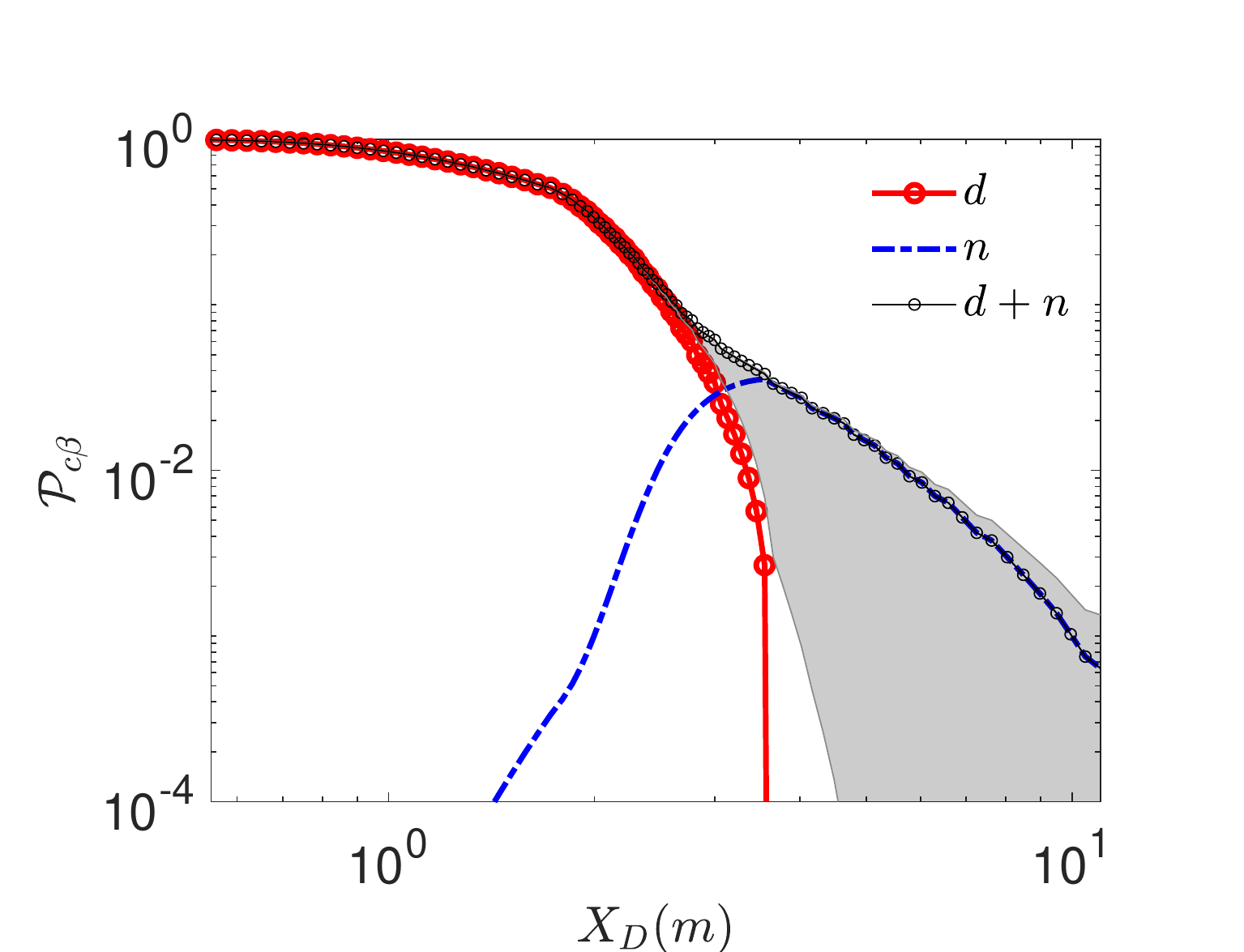}
\caption{}
\label{Fig:Pvd_Pvn(b)}
\end{subfigure}
\caption{Probability of infection $\mathcal{P}_{c \beta}$ for droplet route $d$ and dried droplet nuclei route $n$, as well as total probability $\mathcal{P}_{\beta}$  for (a) as a function time measured from the instant of the beginning of the expiratory event (b) as a function of distance measured from location of the origin of the expiratory event along the center of the jet/puff trajectory. $T_{\infty}=21.44^o C, \ RH_{\infty}=50 \%$. The bold lines represent $\rho_v = 7 \times 10^6$ copies/ml with $t_{d\frac{1}{2}} = t_{n\frac{1}{2}}=t_{\frac{1}{2}}$, where $t_{\frac{1}{2}}=15.25$ minutes. The grey shaded region denotes the lower limit $t_{n\frac{1}{2}} = 0.01t_{d\frac{1}{2}}$ and upper limit $t_{n\frac{1}{2}} = 100t_{d\frac{1}{2}}$, respectively, with $t_{d\frac{1}{2}} = t_{\frac{1}{2}}$.}
\label{Fig:Pvd_Pvn}
\end{centering}
\end{figure}

\begin{figure}[ht!]
\begin{centering}
\begin{subfigure}{0.470\textwidth}
\includegraphics[trim=3cm 8.5cm 3cm 7.5cm,clip,width=\textwidth]{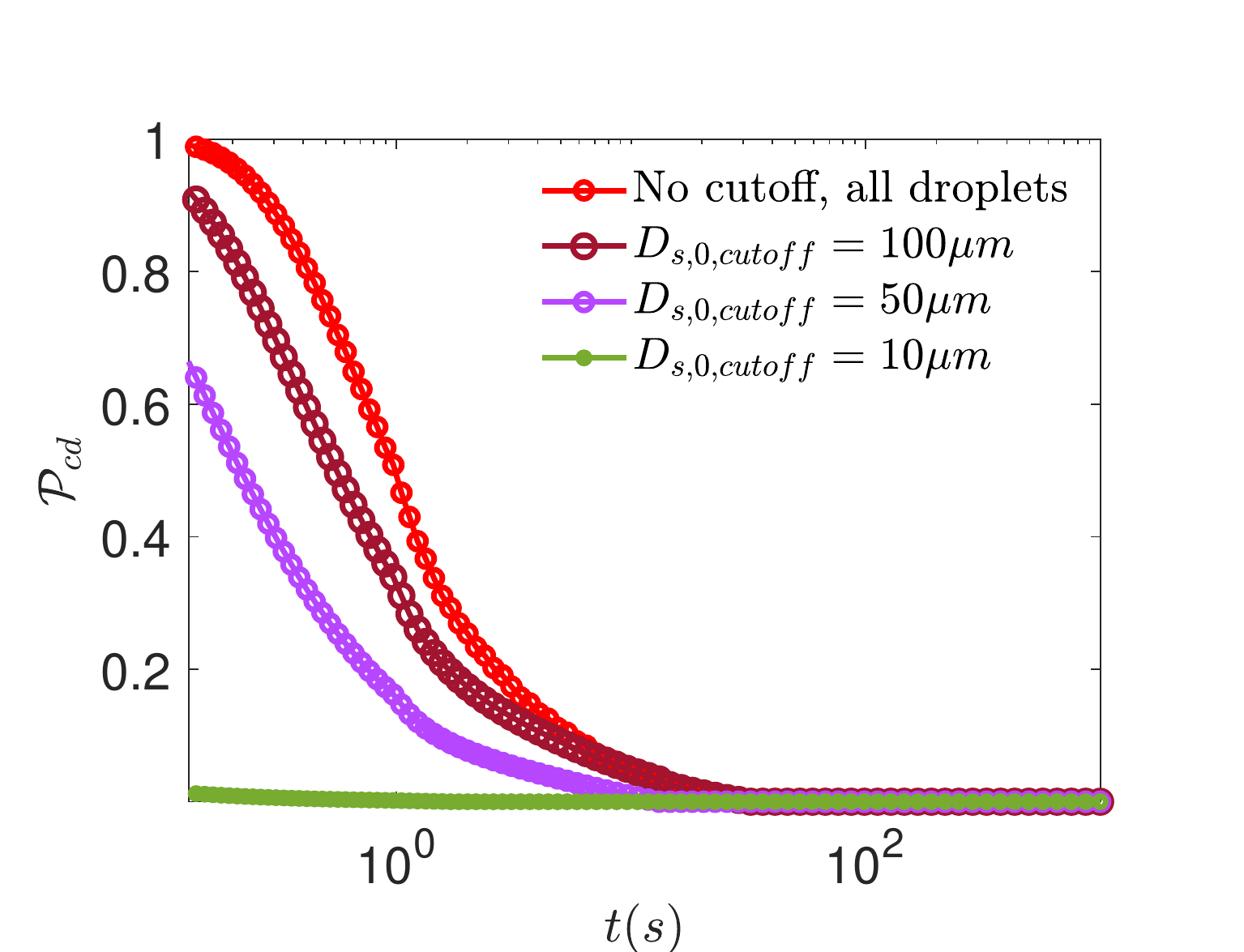}
\caption{}
\label{Fig:Pvd(a)}
\end{subfigure}
\begin{subfigure}{0.470\textwidth}
\includegraphics[trim=3cm 8.5cm 3cm 7.5cm,clip,width=\textwidth]{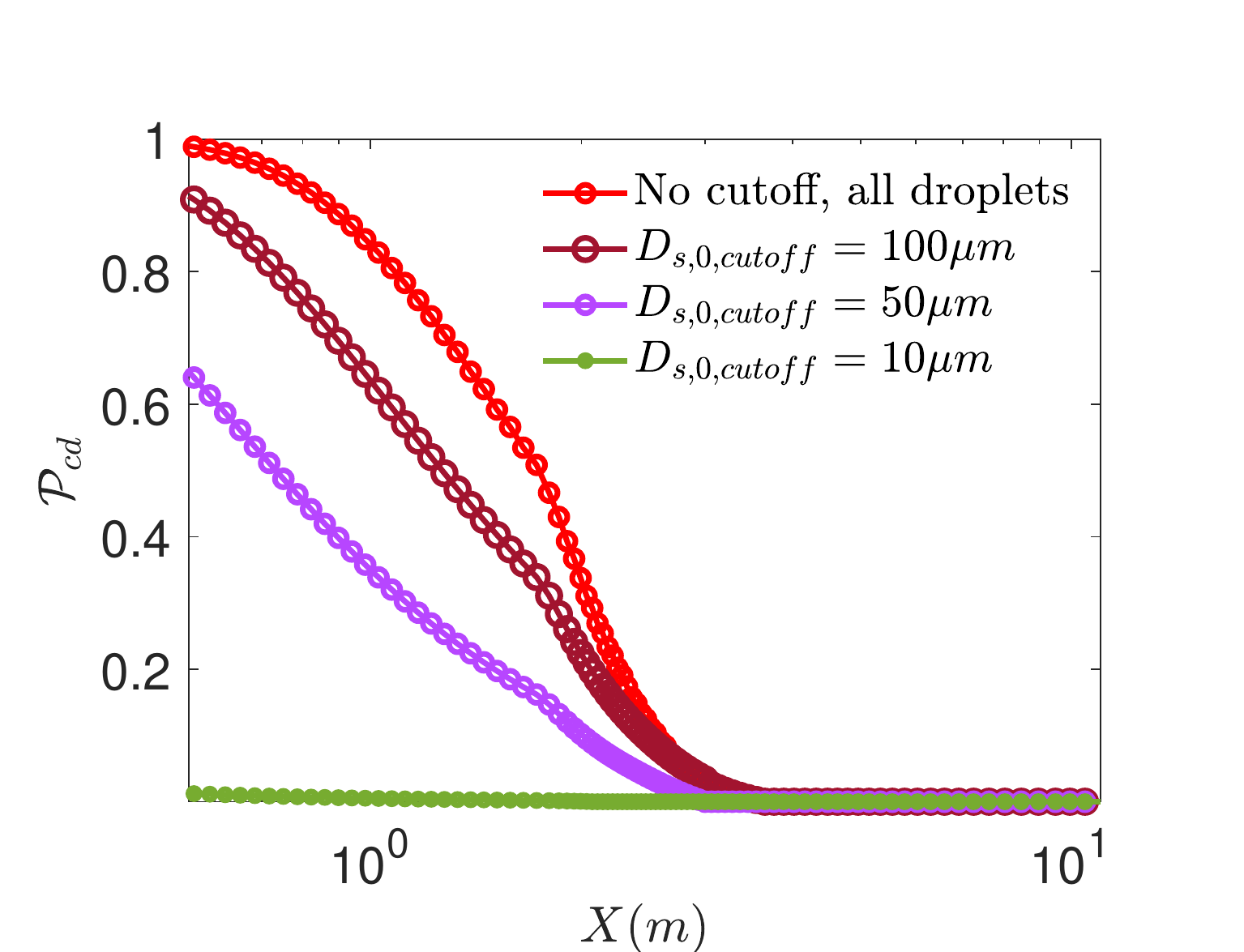}
\caption{}
\label{Fig:Pvd(b)}
\end{subfigure}
\caption{Probability of infection $\mathcal{P}_{cd}$ for droplet route $d$ at different cutoff droplet sizes (implying no droplets beyond that size) for (a) as a function of time measured from the instant of the beginning of the expiratory event (b) as a function of distance measured from location of the origin of the expiratory event along the center of the jet/puff trajectory. $T_{\infty}=21.44^o C, \ RH_{\infty}=50 \%$. $\rho_v = 7 \times 10^6$ copies/ml with $t_{d\frac{1}{2}} = t_{n\frac{1}{2}}=t_{\frac{1}{2}}$, where $t_{\frac{1}{2}}=15.25$ minutes.}
\label{Fig:Pvd}
\end{centering}
\end{figure}

\begin{figure}[hbt!]
\begin{centering}
\begin{subfigure}{0.470\textwidth}
\includegraphics[trim=3cm 8.5cm 3cm 7.5cm,clip,width=\textwidth]{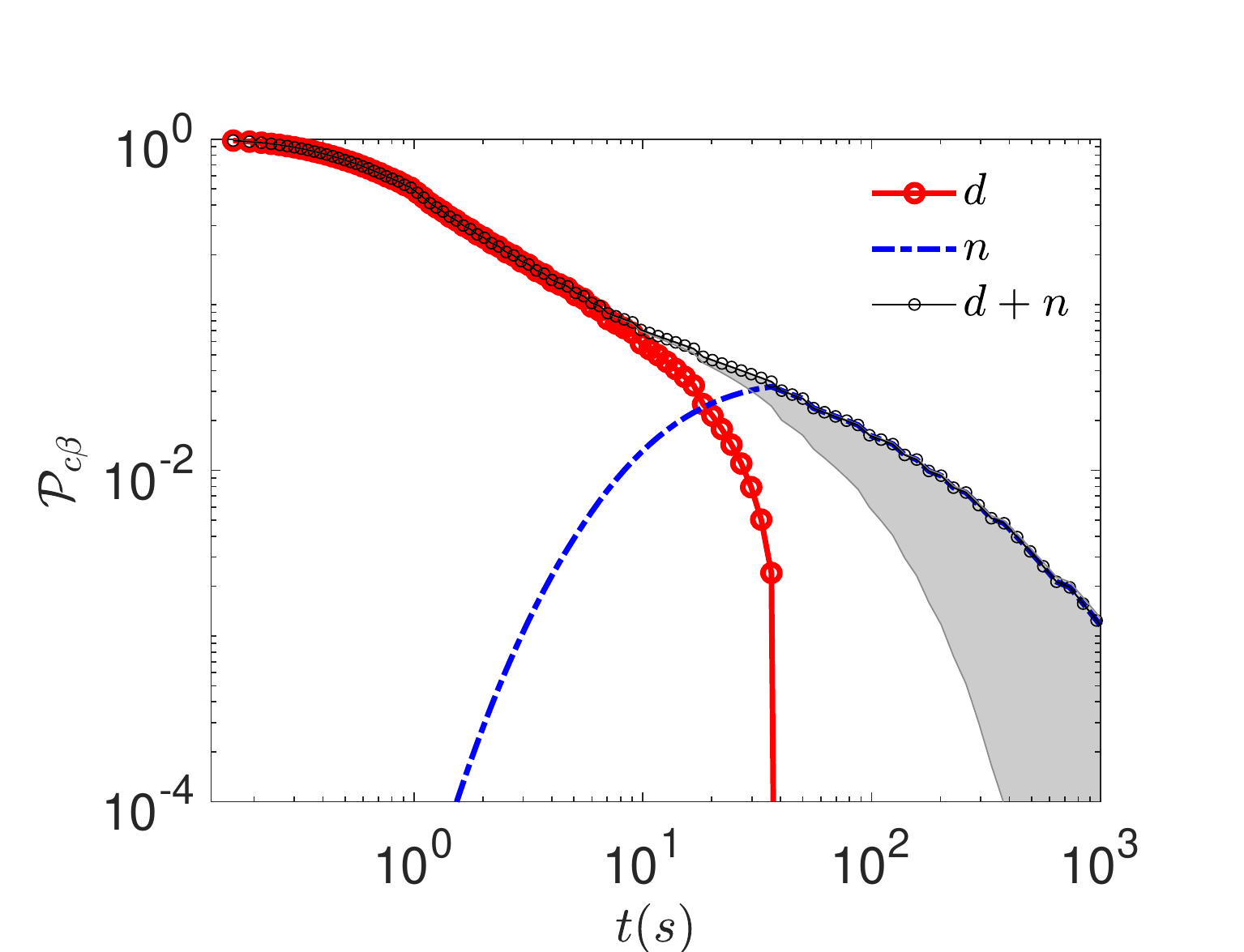}
\caption{}
\label{Fig:Pvd_Pvn(a)_20}
\end{subfigure}
\begin{subfigure}{0.470\textwidth}
\includegraphics[trim=3cm 8.5cm 3cm 7.5cm,clip,width=\textwidth]{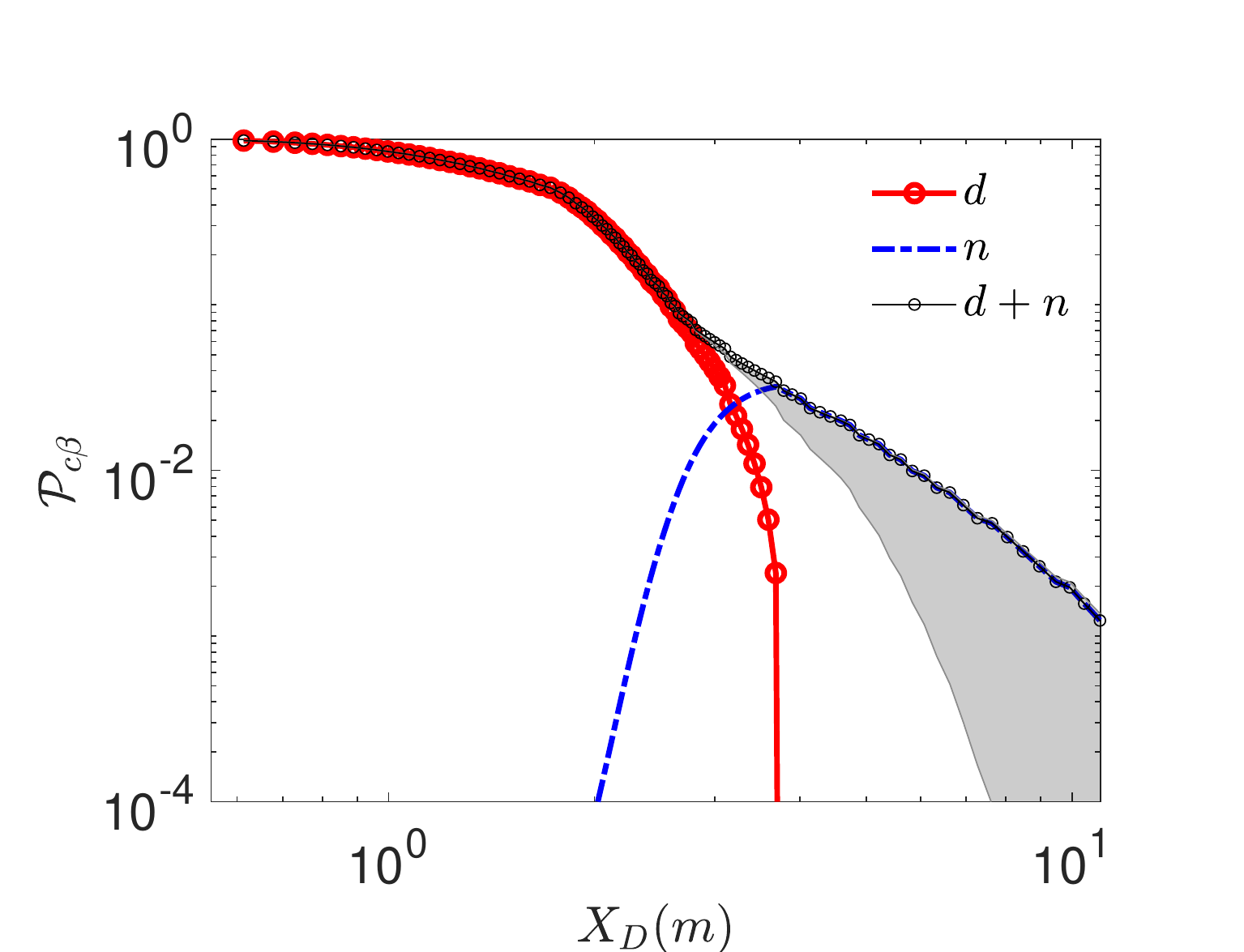}
\caption{}
\label{Fig:Pvd_Pvn(b)_20}
\end{subfigure}
\caption{Probability of infection $\mathcal{P}_{c \beta}$ for droplet route $d$ and dried droplet nuclei route $n$ (a) as a function time measured from the instant of the beginning of the expiratory event (b) as a function of distance measured from location of the origin of the expiratory event along the center of the jet/puff trajectory. $T_{\infty}=10^o C, \ RH_{\infty}=20 \%$ and $\rho_v = 7 \times 10^6$ copies/ml.}
\label{Fig:Pvd_Pvn_20}
\end{centering}
\end{figure}

Next, we analyze the time varying infection probability. Interestingly in Fig. \ref{Fig:Pvd_Pvn(a)}, at $T_{\infty}=21.44^o C, \ RH_{\infty}=50 \%$ for $t>1s$ the total probability of infection scales as $\mathcal{P}_{c} \sim t^{-2/3}$, for droplets and dried droplet nuclei. This is a combined effect of droplet evaporation, virus decay and dilution due to entrainment of fresh air within the jet/puff, the diameter of which increases initially as $t^{1/2}$ and then as $t^{1/4}$, respectively. After the droplets evaporate, the decay of the infection probability for the dried droplet nuclei slows down with respect to their droplet predecessors. This is because, while the infection probability decay for droplets is due to evaporation, settling and dilution, the probability decay due to nuclei is due to only dilution and finite virus lifetime. Here, we are considering a very large, poorly ventilated indoor space with a large number of occupants as in a shopping mall or in a conference center. It is to be noted that the dilution effect is arrived with the assumption that all people are in motion but their motion do not affect the cloud aerodynamics. In reality, such motion will lead to increased turbulence and mixing, resulting in further dilution.  Therefore, the decay of the probability of infection will be faster than $ t^{-2/3}$ in reality. Another interesting feature is that the $\mathcal{P}_{cn}$ does not continuously follow the $\mathcal{P}_{cd}$ after all the droplets evaporate. There is an accumulation of the droplet nuclei due to evaporation of smaller droplets beforehand leading to a small jump in $\mathcal{P}_{cn}$ at $\tau_d$. Indeed the overall probability $\mathcal{P}_c$ by Eqn. \ref{Eq:Pv sum equation} decreases without any discontinuity. From Fig. \ref{Fig:Pvd_Pvn(b)} we find that for $X_D>2m$ the overall probability decreases as $X_D^{-3}$ justifying the necessity of social distancing. However, $\mathcal{P}_c < 0.01$ only after about 5m.

The effect of preventing ejection of droplets beyond particular initial sizes, on $\mathcal{P}_{cd}$ is examined in Fig. \ref{Fig:Pvd(a)}, \ref{Fig:Pvd(b)} as a function of time and distance. For $D_{s,0,cutoff}=50 \mu m$ an infection probability of 0.6 is obtained for $t \rightarrow{} 0$ suggesting droplets with $D_{s,0}<50 \mu m$ is slightly more responsible for infection, at all times, for the conditions under consideration, than their $D_{s,0}>50 \mu m$ counterparts. This trend continues until $\tau_d$ when all airborne droplets evaporate. This is qualitatively consistent with the exposure analysis and results of Chen et al. \cite{chen2020short} who considered dispersion and evaporation of water droplets with size distribution from Duguid \cite{duguid1946size}. However, when $D_{s,0,cutoff}=10 \mu m$, the corresponding probability of infection is very small, suggesting that for the average viral loading, at early times the droplets of initial diameter $10 \mu m<D_{s,0}<50 \mu m$ are the most lethal in terms of their probability to infect. However, while they infect their diameters are substantially smaller.

The $\mathcal{P}_{c\beta}$ at $T_{\infty}=10^o C, \ RH_{\infty}=20 \%$ is shown in Fig. \ref{Fig:Pvd_Pvn(a)_20}, \ref{Fig:Pvd_Pvn(b)_20} as a function of time and distance. In comparison to the previous case, here, the droplet survives longer due to lower temperature while droplet-nuclei induces higher $\mathcal{P}_{cn}$ due to longer virus half-life. Thus at lower temperature, higher infection probability could be expected.
However, in both cases, at short time and distance from the expiratory event, droplets (both small and large) dominate transmission and only after most droplets evaporate, the dried droplet nuclei route is significantly activated. While the transmission probability by dried droplet nuclei are always lower than by droplets, their lifetime is theoretically infinite as opposed to the finite lifetime $\tau_d$ of the droplets. Hence, despite their low instantaneous probability of infection, cumulatively they contribute significantly, assuming that the virus remain infectious for significant times  within the dried droplet nuclei. If so, as will be shown below, the persistent dried droplet nuclei appears to be a major transmission mode of the virus. It is to be recognized that these results were obtained with average viral loading $\rho_v = 7 \times 10^6$ copies/ml. If we consider the $\rho_{v,max} = 2.35 \times 10^9$ copies/ml, $\mathcal{P}_{c \beta}$ does not decay from the maximum fixed value of $1$ until from about $100s$ or from $5m$ from the origin of the respiratory jet, along the center of the jet. Thus it is expected that such kind of viral loading could infect a large number of $S$ potentially leading to a superspreading event. 

Using the $\mathcal{P}_{\alpha \beta}(t)$ thus obtained, we can evaluate the corresponding rate constants for $d$ and $n$ using Eqn. \ref{Eq:k1_alpha_d equation} and Eqn. \ref{Eq:k1_alpha_n equation}, respectively. These rate constants for $\rho_v = 7 \times 10^6$ copies/ml are presented in Table \ref{tab:2} for four Cases IA, IB, IC and II.  Cases I(A-C) corresponds to $T_{\infty}=21.44^o C, \ RH_{\infty}=50 \%$ while Case II represents $T_{\infty}=10^o C, \ RH_{\infty}=20 \%$. Population density in both cases is assumed to be $10000$ people/km$^2$. In all cases, homogeneous mixing is assumed without any social distancing or lockdown. In all cases $k_{1,cn\gamma}>k_{1,cd\gamma}$. Case IA represents no restriction and clearly high rate constant values are attained in this case. Case IB represents a hypothetical situation where the ejection of all droplets with $D_{s,0}>10 \mu m$ is restricted. This is hypothetically possible by stringent enforcement of population wide usage of ordinary face-masks without any exceptions. Furthermore, using $k_{1,\alpha}$ and $k_3$ we can define the basic reproduction number $\mathcal{R}_{0,
\alpha}=k_{1,\alpha}/(0.97 k_3)$. The calculated $k_{1,c\beta}$ and $\mathcal{R}_{0,c}$ could be found in Table \ref{tab:2}. A very interesting $\mathcal{R}_{0,c}$ trend emerges between Case IA and IB. We find that if ejection of droplets even beyond $10 \mu m$ could be completely prevented, the $\mathcal{R}_{0,c}$ drops from $4.22 \ (0.33, 5.63)$ for Case IA to $0.048$ for Case IB. For Case 1A, the numbers in the brackets denote the $\mathcal{R}_{0,c}$ for lower limit $t_{n\frac{1}{2}} = 0.01t_{d\frac{1}{2}}$ and upper limit $t_{n\frac{1}{2}} = 100t_{d\frac{1}{2}}$, respectively. The $\mathcal{R}_{0,c}$ between Case 1A and 1B represent two order of magnitude difference and for the $\mathcal{R}_{0,c}$ at Case IB, no outbreak is possible. The bifurcation point $\mathcal{R}_{0,c} \approx 1$ is attained for the critical droplet size $D_{s,0}=27 \mu m$. This is shown in Table \ref{tab:2} as Case IC. The implication is that preventing ejection of droplets with initial size beyond $27 \mu m$ would just prevent the outbreak. Of course, it is to be recognized that we are only considering cough as the mode of droplet ejection alongside many idealizing assumptions. Furthermore, these results were arrived at with the average viral load $\rho_v=7 \times 10^6$ copies/ml. If we consider the maximum reported viral load $\rho_{v,max}= 2.35 \times 10^9$ copies/ml with free mixing among $I$ and $S$, the $\mathcal{R}_{0,c}=634.12 \ (16.32, 869.28)$,  indicating a superspreading event. As such, it could be a combination of high mobility and large viral loading of $I$ -  that could lead to a super-spreader. 

\begin{figure}[hbt!]
\includegraphics[trim=1cm 8.5cm 1cm 7.5cm,clip,width=1\textwidth]{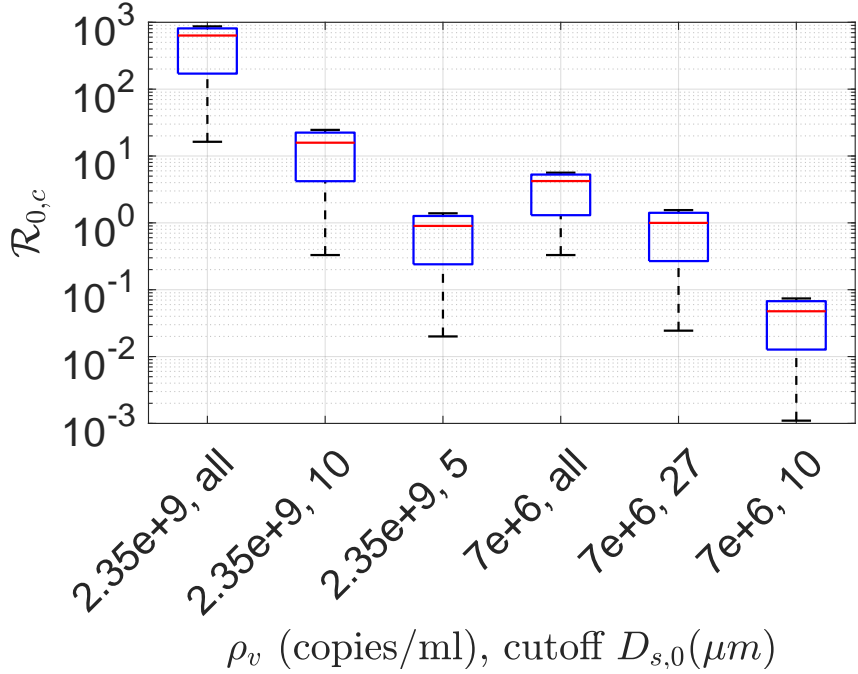}
\begin{centering}
\caption{Comparison of $\mathcal{R}_{0,c}$ for different conditions based on two different viral loading $\rho_{v,maximum}=2.35 \times 10^9$ copies/ml, $\rho_{v,average}=7 \times 10^6$ copies/ml and if ejection of droplets beyond the specified cutoff sizes are prevented. In each boxes the red-line denotes the equal half-life condition $t_{n\frac{1}{2}}=t_{d\frac{1}{2}}$, irrespective of the phase. The lower and upper limit corresponds to $t_{n\frac{1}{2}}=0.01t_{d\frac{1}{2}}$ and $t_{n\frac{1}{2}}=100t_{d\frac{1}{2}}$, respectively. All data are obtained at $T_{\infty}=21.44^o C, \ RH_{\infty}=50 \%$ and with $r_v=0.5$ implying a minimum infectious dose of $10$ virions.}
\label{Fig:R0_boxplot}
\end{centering}
\end{figure}

\begin{figure}[hbt!]
\begin{centering}
\begin{subfigure}{0.470\textwidth}
\includegraphics[trim=3cm 8.5cm 3cm 7.5cm,clip,width=\textwidth]{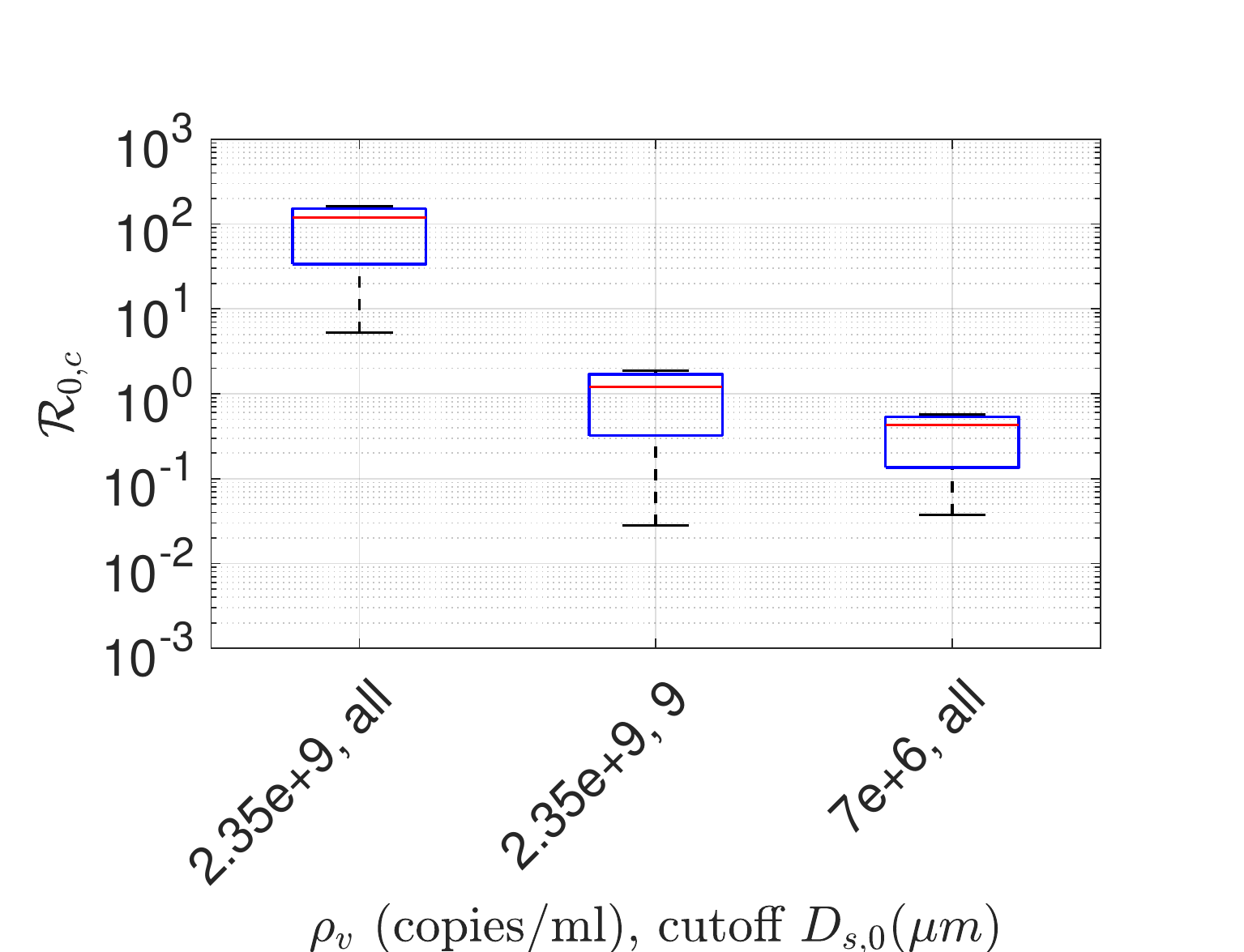}
\caption{$r_v=0.05$}
\label{Fig:R0Boxplot_rv005}
\end{subfigure}
\begin{subfigure}{0.470\textwidth}
\includegraphics[trim=3cm 8.5cm 3cm 7.5cm,clip,width=\textwidth]{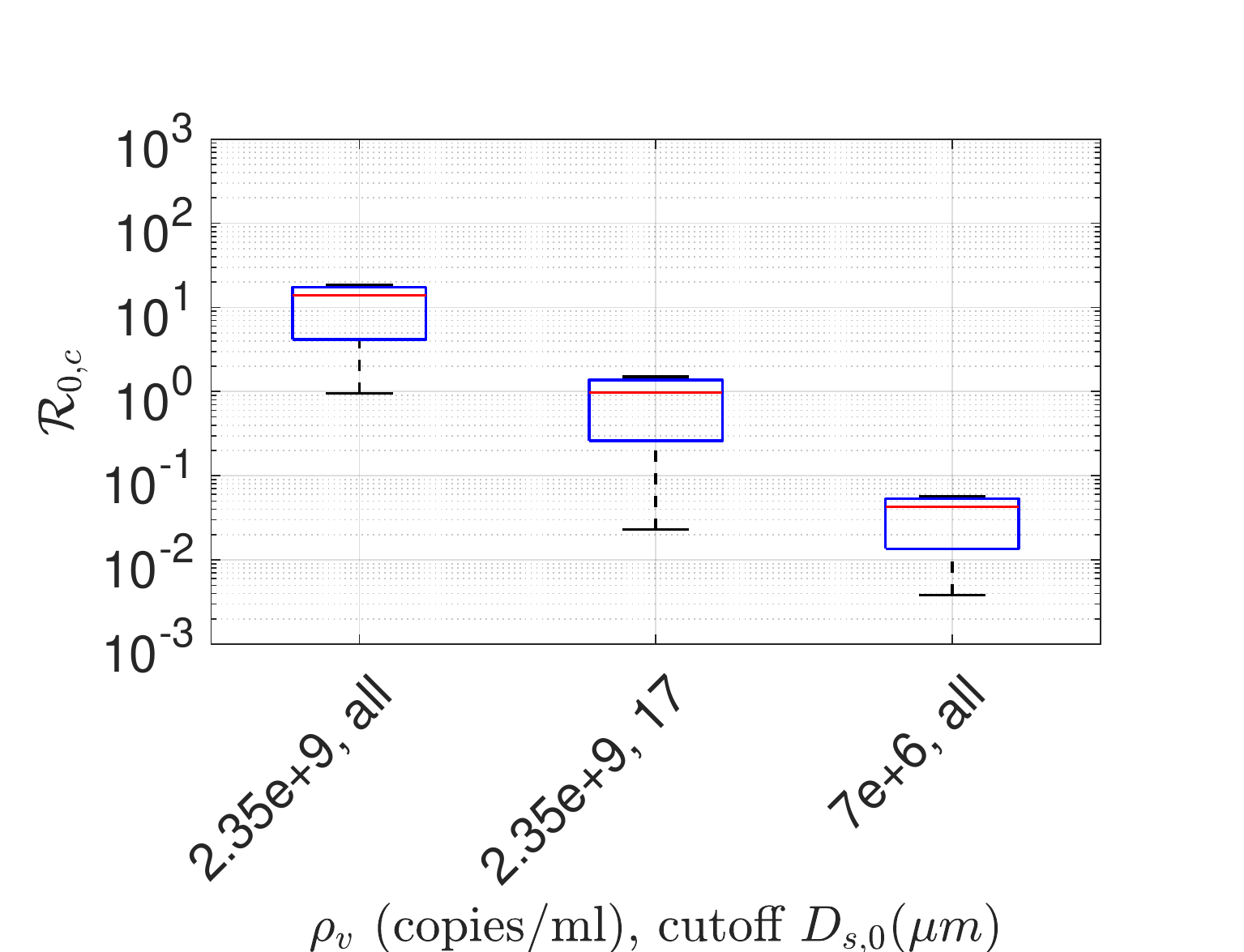}
\caption{$r_v=0.005$}
\label{Fig:R0Boxplot_rv0005}
\end{subfigure}
\caption{Comparison of $\mathcal{R}_{0,c}$ for different conditions based on two different viral loading $\rho_{v,maximum}=2.35 \times 10^9$ copies/ml, $\rho_{v,average}=7 \times 10^6$ copies/ml and if ejection of droplets beyond the specified cutoff sizes are prevented. In each boxes the red-line denotes the equal half-life condition $t_{n\frac{1}{2}}=t_{d\frac{1}{2}}$, irrespective of the phase. The lower and upper limit corresponds to $t_{n\frac{1}{2}}=0.01t_{d\frac{1}{2}}$ and $t_{n\frac{1}{2}}=100t_{d\frac{1}{2}}$, respectively. All data are obtained at $T_{\infty}=21.44^o C, \ RH_{\infty}=50 \%$ and with (a) $r_v=0.05$ implying a minimum infectious dose of $100$ virions (b) $r_v=0.005$ implying a minimum infectious dose of $1000$ virions.}
\label{Fig:R0_Boxplot_rv005_rv_0.005}
\end{centering}
\end{figure}

$\mathcal{R}_{0,c}$ calculated at the average and maximum viral loading $\rho_v=7 \times 10^6$ copies/ml and $\rho_{v,max}=2.35 \times 10^9$ copies/ml, respectively, for different droplet size cutoffs at $T_{\infty}=21.44^o C, \ RH_{\infty}=50 \%$ is shown in Fig. \ref{Fig:R0_boxplot}. The cutoff $D_{s,0}$ means all droplets with sizes $D_{s,0}>D_{s,0,cutoff}$ are prevented from ejecting. This figure also shows the sensitivity of the assumption $t_{n\frac{1}{2}}=t_{d\frac{1}{2}}=t_{\frac{1}{2}}$ on the results. In Fig. \ref{Fig:R0_boxplot}, the lower and upper limits represent the conditions $t_{n\frac{1}{2}}=0.01t_{d\frac{1}{2}}$ and $t_{n\frac{1}{2}}=100t_{d\frac{1}{2}}$, respectively. If all droplets are allowed to be ejected at average viral loading, for  $t_{n\frac{1}{2}}=0.01t_{d\frac{1}{2}}$, $\mathcal{R}_{0,c}=0.33$ while for $t_{n\frac{1}{2}}=100t_{d\frac{1}{2}}$, $\mathcal{R}_{0,c}=5.63$ with the base $\mathcal{R}_{0,c}=4.22$ for the typical indoor conditions assumed above. Clearly, change in the lower limit of $t_{n\frac{1}{2}}$ is much more sensitive than its upper limit. This is because even if the viral lifetime is much longer, dilution reduces infection probability. However, with $\rho_{v,max}= 2.35 \times 10^9$ copies/ml for $t_{n\frac{1}{2}}=0.01t_{d\frac{1}{2}}$, $\mathcal{R}_{0,c}=16.32$ while for $t_{n\frac{1}{2}}=100t_{d\frac{1}{2}}$, $\mathcal{R}_{0,c}=869.28$ around the base case of $\mathcal{R}_{0,c}=634.12$ all other conditions remaining same. Interestingly, with a $D_{s,0,cutoff}=10 \mu m$ the $\mathcal{R}_{0,c}$ reduces by a factor of 40 w.r.t. no cutoff condition. However, $D_{s,0,cutoff}=5 \mu m$ reduces the $\mathcal{R}_{0,c}$ by another factor of 18 w.r.t. $D_{s,0,cutoff}=10 \mu m$ condition. for the base cases. At this condition $\mathcal{R}_{0,c}<1$. For both viral loading, the maximum and the averaged, blocking droplets $D_{s,0} \geq 5 \mu m$ can theoretically yield a $\mathcal{R}_{0,c} \leq 1$. All the results so far, have been obtained with $r_v=0.5$ which implies a minimum infectious dose of $10$ virions. Figures \ref{Fig:R0_Boxplot_rv005_rv_0.005}(a) and (b) shows the corresponding $\mathcal{R}_{0,c}$ for $r_v=0.05$
and $r_v=0.005$, respectively. These imply minimum infectious doses of 100 and 1000 virions respectively. While the qualitative trend is similar, indeed the $\mathcal{R}_{0,c}$ for these two cases are much lower in comparison to $r_v=0.5$. As such it seems likely that the minimum infectious dose of SARS-Cov-2 is $\mathcal{O}(10)$. While the base $\mathcal{R}_{0,c}=4.22$ for the average viral loading, $r_v=0.5$ and no cutoff is consistent with that of reported values for Covid-19 \cite{liu2020reproductive}, the order of magnitude larger values of $\mathcal{R}_{0,c}$ obtained at maximum viral loading should be viewed in context of superspreading events.

\begin{figure}[ht!]
\begin{centering}
\begin{subfigure}{0.470\textwidth}
\includegraphics[trim=3cm 8.5cm 3cm 7.5cm,clip,width=\textwidth]{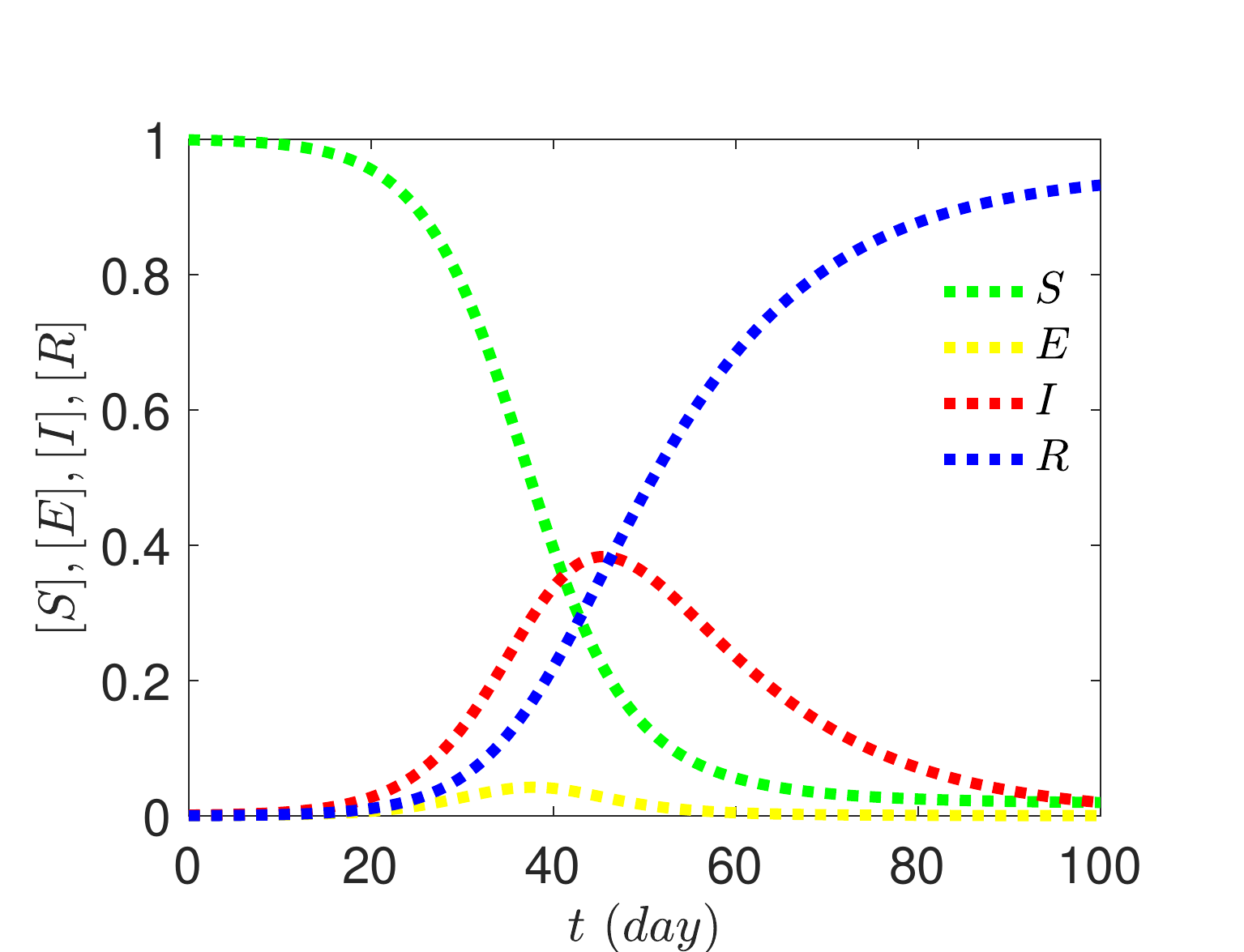}
\caption{}
\label{Fig:SEIR(a)}
\end{subfigure}
\begin{subfigure}{0.470\textwidth}
\includegraphics[trim=3cm 8.5cm 3cm 7.5cm,clip,width=\textwidth]{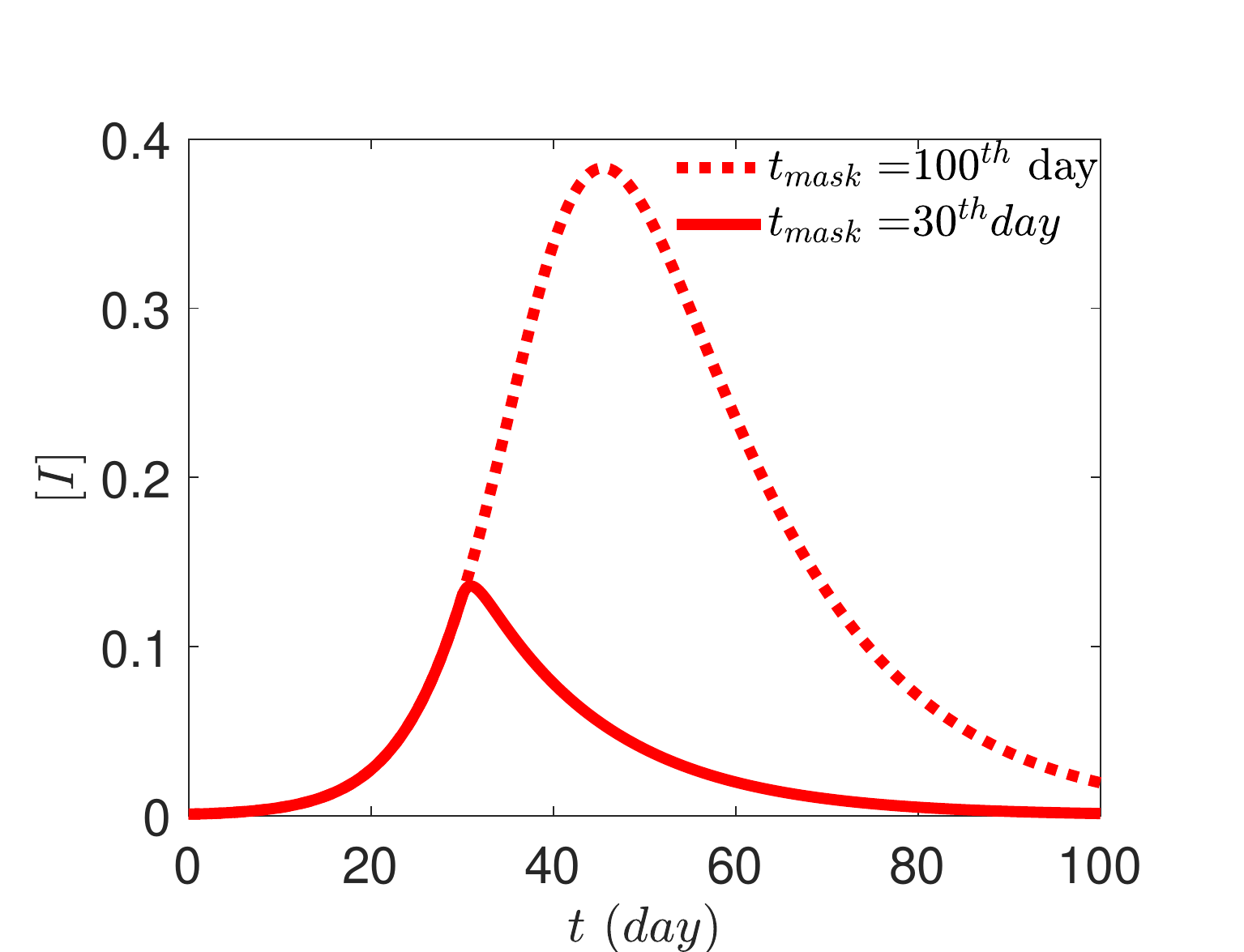}
\caption{}
\label{Fig:SEIR(b)}
\end{subfigure}
\caption{SEIR-droplet/nuclei model output with computed rate constants mentioned in Table \ref{tab:2} for Case IA. No lock-down and free mixing is assumed (a) Case IA, no mask usage (b) Case IB, all $I$ wear masks from i) $100^{th}$ ii) $30^{th}$ day of the onset of the pandemic that prevents ejection of $D_{s,0}>10 \mu m$ droplets with $\rho_v=7 \times 10^6$ copies/ml. Calculated rate constants are shown in Table \ref{tab:2} for Case IA and IB. The masks for this case are assumed to prevent ejection of all particles greater than $10 \mu m$.}
\label{Fig:Pvd_Pvn}
\end{centering}
\end{figure}

Using the governing Eqns. \ref{Eq:SEIR_ODE} the evolution of the pandemic for average viral loading and $r_v=0.5$, for Case IA is presented in Fig. \ref{Fig:SEIR(a)}. The growth rate of the infected population for Case IA and Case IB is shown in Fig. \ref{Fig:SEIR(b)} with the assumption that usage of face masks for the entire $I$ population (which would  practically be required for the entire population) is implemented after a fixed time from the onset of the outbreak. As expected, in this SEIR model with \textit{ab initio} infection rate constants, the effect of the usage of masks is almost immediate since the assumed latency period is only 1 day. Leffler et al. \cite{Leffler2020.05.22.20109231} analyzed Covid-19 data from 198 countries to conclude that government policies on mask wearing significantly reduced mortality.  A significant feature of the results of this paper is that, while they are computed mechanistically from first principles with assumptions and limitations, they produce physically meaningful outcomes.

\begin{table}[h!]
\begin{center}
\begin{tabular}{ | m{1cm} | m{7cm} | m{2cm}| m{2cm} | m{2cm} |  } 
\hline
Case & Condition & $k_{1,cd \gamma}$ & $k_{1,cn \gamma}$ & $R_{0,c}$\\ 
\hline
IA & cough, no mask &  0.0182
 & 0.2743
 &  4.2219 \\
\hline
IB& cough, $D_{s,0}=10 \mu m$ cutoff mask for all  &  1.58e-05
&  0.0033 & 0.0476 \\
\hline
IC& cough, $D_{s.0}=27 \mu m$ cutoff mask for all  &  6.23e-04
&  0.0686 &  0.9981\\
\hline
II & cough, no mask  &  0.0229
&  0.3519
 &  5.4087
 \\
\hline
\end{tabular}
\caption{\label{tab:2} Calculated infection rate constant values for different modes of transmission for coughing with and without mask at typical indoor conditions. Case IA, IB and IC: $T_{\infty}=21.1^o C, \ RH_{\infty}=50 \%$. Case II: $T_{\infty}=10^o C, \ RH_{\infty}=20 \%$. For all cases $\rho_v=7 \times 10^6$ copies/ml.}
\end{center}
\end{table}

\section{Conclusions}
We have analyzed the relative significance of the different transmission modes of SARS-CoV-2, using first principle calculations. Starting with a well known cough droplet size distribution, we derived the time dependent probability of infection by different routes accounting for viral load, virus stability, respiratory droplet cloud aerodynamics, evaporation and crystallization for poorly ventilated conditions. Most number of droplets have an initial diameter of $D_{s,0}=13.9 \mu m$, but within $1s$ of their ejection, most number of droplets of the same set gets reduced to a diameter of $2.7\mu m$ due to evaporation, accounting for the thermodynamic state of the exhaled air. For the average viral loading, at early times, the droplets of initial diameter $10 \mu m<D_{s,0}<50 \mu m$ are the most lethal in terms of their probability to infect. However, while they infect, their diameters could be $5-6$ times smaller. Indeed for most of the time, infection is spread by inhalation of small, airborne droplets or their desiccated nuclei. While the instantaneous probability of infection by droplets is significantly larger than its dried nuclei in the short time and range, the much longer persistence of the dried nuclei results in its stronger relative contribution to the infection rate constant, under the assumption that the virus half-life is independent of the phase of its vector. The infection rate constant is derived \textit{ab initio} by calculating collision frequency between the droplets/nuclei cloud and the susceptible population for different ambient conditions including the probability of infection. The SEIR model output obtained with the calculated rate constants for average viral loading, for the specific conditions of interest, show that preventing ejection of droplets with initial diameter greater than $10 \mu m$ can potentially prevent further outbreaks even for a minimum infectious dose of 10 virions. The critical droplet diameter, preventing ejection of droplets above which would result in $\mathcal{R}_{0,c} \approx 1$ is found to be $27 \mu m$. For maximum viral loading, the critical droplet diameter is $5 \mu m$ to just prevent the outbreaks. Furthermore, strong sensitivity of $\mathcal{R}_{0,c}$ on variation of virus half-life at different phases of the droplet/aerosol, as well as the minimum infectious dose is demonstrated. 
\section{Acknowledgements}
The authors thank Prof. S. Balachandar from University of Florida, for his comments on the manuscript.

\bibliography{bibliography}{}

\providecommand{\noopsort}[1]{}\providecommand{\singleletter}[1]{#1}%
\begin{thebibliography}{10}

\bibitem{Morawska2020Commentary}
L.~Morawska and D.~K. Milton, ``{It is Time to Address Airborne Transmission of
  COVID-19},'' {\em Clinical Infectious Diseases}, 07 2020.
\newblock ciaa939.

\bibitem{liu2020aerodynamic}
Y.~Liu, Z.~Ning, Y.~Chen, M.~Guo, Y.~Liu, N.~K. Gali, L.~Sun, Y.~Duan, J.~Cai,
  D.~Westerdahl, {\em et~al.}, ``Aerodynamic analysis of sars-cov-2 in two
  wuhan hospitals,'' {\em Nature}, pp.~1--4, 2020.

\bibitem{WHOJuly92020}
W.~H. Organization {\em et~al.}, ``Transmission of sars-cov-2: implications for
  infection prevention precautions, scientific brief. 9 july 2020,'' tech.
  rep., World Health Organization, 2020.

\bibitem{mittal2020flow}
R.~Mittal, R.~Ni, and J.-H. Seo, ``The flow physics of covid-19,'' {\em Journal
  of fluid Mechanics}, vol.~894, 2020.

\bibitem{stadnytskyi2020airborne}
V.~Stadnytskyi, C.~E. Bax, A.~Bax, and P.~Anfinrud, ``The airborne lifetime of
  small speech droplets and their potential importance in sars-cov-2
  transmission,'' {\em Proceedings of the National Academy of Sciences},
  vol.~117, no.~22, pp.~11875--11877, 2020.

\bibitem{duguid1945numbers}
J.~Duguid, ``The numbers and the sites of origin of the droplets expelled
  during expiratory activities,'' {\em Edinburgh Medical Journal}, vol.~52,
  no.~11, p.~385, 1945.

\bibitem{xie2009exhaled}
X.~Xie, Y.~Li, H.~Sun, and L.~Liu, ``Exhaled droplets due to talking and
  coughing,'' {\em Journal of the Royal Society Interface}, vol.~6,
  no.~suppl\_6, pp.~S703--S714, 2009.

\bibitem{chao2009characterization}
C.~Y.~H. Chao, M.~P. Wan, L.~Morawska, G.~R. Johnson, Z.~Ristovski,
  M.~Hargreaves, K.~Mengersen, S.~Corbett, Y.~Li, X.~Xie, {\em et~al.},
  ``Characterization of expiration air jets and droplet size distributions
  immediately at the mouth opening,'' {\em Journal of Aerosol Science},
  vol.~40, no.~2, pp.~122--133, 2009.

\bibitem{bourouiba2014violent}
L.~Bourouiba, E.~Dehandschoewercker, and J.~W. Bush, ``Violent expiratory
  events: on coughing and sneezing,'' {\em Journal of Fluid Mechanics},
  vol.~745, pp.~537--563, 2014.

\bibitem{bourouiba2020turbulent}
L.~Bourouiba, ``Turbulent gas clouds and respiratory pathogen emissions:
  potential implications for reducing transmission of covid-19,'' {\em Jama},
  vol.~323, no.~18, pp.~1837--1838, 2020.

\bibitem{vejerano2018physico}
E.~P. Vejerano and L.~C. Marr, ``Physico-chemical characteristics of
  evaporating respiratory fluid droplets,'' {\em Journal of The Royal Society
  Interface}, vol.~15, no.~139, p.~20170939, 2018.

\bibitem{marr2019mechanistic}
L.~C. Marr, J.~W. Tang, J.~Van~Mullekom, and S.~S. Lakdawala, ``Mechanistic
  insights into the effect of humidity on airborne influenza virus survival,
  transmission and incidence,'' {\em Journal of the Royal Society Interface},
  vol.~16, no.~150, p.~20180298, 2019.

\bibitem{lin2019humidity}
K.~Lin and L.~C. Marr, ``Humidity-dependent decay of viruses, but not bacteria,
  in aerosols and droplets follows disinfection kinetics,'' {\em Environmental
  Science \& Technology}, vol.~54, no.~2, pp.~1024--1032, 2019.

\bibitem{keeling2011modeling}
M.~J. Keeling and P.~Rohani, {\em Modeling infectious diseases in humans and
  animals}.
\newblock Princeton University Press, 2011.

\bibitem{bertozzi2020challenges}
A.~L. Bertozzi, E.~Franco, G.~Mohler, M.~B. Short, and D.~Sledge, ``The
  challenges of modeling and forecasting the spread of covid-19,'' {\em arXiv
  preprint arXiv:2004.04741}, 2020.

\bibitem{adam2020special}
D.~Adam, ``Special report: The simulations driving the world's response to
  covid-19.,'' {\em Nature}, vol.~580, no.~7803, p.~316, 2020.

\bibitem{metcalf2020mathematical}
C.~J.~E. Metcalf, D.~H. Morris, and S.~W. Park, ``Mathematical models to guide
  pandemic response,'' {\em Science}, vol.~369, no.~6502, pp.~368--369, 2020.

\bibitem{wolfel2020virological}
R.~W{\"o}lfel, V.~M. Corman, W.~Guggemos, M.~Seilmaier, S.~Zange, M.~A.
  M{\"u}ller, D.~Niemeyer, T.~C. Jones, P.~Vollmar, C.~Rothe, {\em et~al.},
  ``Virological assessment of hospitalized patients with covid-2019,'' {\em
  Nature}, vol.~581, no.~7809, pp.~465--469, 2020.

\bibitem{barrett2019ganong}
K.~E. Barrett, S.~M. Barman, H.~L. Brooks, and J.~X.-J. Yuan, {\em Ganong's
  review of medical physiology}.
\newblock McGraw-Hill Education, 2019.

\bibitem{schuit2020airborne}
M.~Schuit, S.~Ratnesar-Shumate, J.~Yolitz, G.~Williams, W.~Weaver, B.~Green,
  D.~Miller, M.~Krause, K.~Beck, S.~Wood, {\em et~al.}, ``Airborne sars-cov-2
  is rapidly inactivated by simulated sunlight,'' {\em The Journal of
  Infectious Diseases}, 2020.

\bibitem{chaudhuri2020modeling}
S.~Chaudhuri, S.~Basu, P.~Kabi, V.~R. Unni, and A.~Saha, ``Modeling the role of
  respiratory droplets in covid-19 type pandemics,'' {\em Physics of Fluids},
  vol.~32, no.~6, p.~063309, 2020.

\bibitem{WHOJune292020}
W.~H. Organization, ``Infection prevention and control during health care when
  coronavirus disease (covid-19) is suspected or confirmed: 29 june 2020,''
  Tech. Rep. WHO/2019-nCoV/IPC/2020.4, World Health Organization, 2020.

\bibitem{haas1983estimation}
C.~N. HAAS, ``Estimation of risk due to low doses of microorganisms: a
  comparison of alternative methodologies,'' {\em American journal of
  epidemiology}, vol.~118, no.~4, pp.~573--582, 1983.

\bibitem{nicas1996analytical}
M.~Nicas, ``An analytical framework for relating dose, risk, and incidence: an
  application to occupational tuberculosis infection,'' {\em Risk Analysis},
  vol.~16, no.~4, pp.~527--538, 1996.

\bibitem{sze2008methodology}
G.~N. Sze~To, M.~Wan, C.~Y.~H. Chao, F.~Wei, S.~Yu, and J.~Kwan, ``A
  methodology for estimating airborne virus exposures in indoor environments
  using the spatial distribution of expiratory aerosols and virus viability
  characteristics,'' {\em Indoor air}, vol.~18, no.~5, pp.~425--438, 2008.

\bibitem{riley1978airborne}
E.~Riley, G.~Murphy, and R.~Riley, ``Airborne spread of measles in a suburban
  elementary school,'' {\em American journal of epidemiology}, vol.~107, no.~5,
  pp.~421--432, 1978.

\bibitem{buonanno2020estimation}
G.~Buonanno, L.~Stabile, and L.~Morawska, ``Estimation of airborne viral
  emission: quanta emission rate of sars-cov-2 for infection risk assessment,''
  {\em Environment International}, p.~105794, 2020.

\bibitem{zwart2009experimental}
M.~P. Zwart, L.~Hemerik, J.~S. Cory, J.~A.~G. de~Visser, F.~J. Bianchi, M.~M.
  Van~Oers, J.~M. Vlak, R.~F. Hoekstra, and W.~Van~der Werf, ``An experimental
  test of the independent action hypothesis in virus--insect pathosystems,''
  {\em Proceedings of the Royal Society B: Biological Sciences}, vol.~276,
  no.~1665, pp.~2233--2242, 2009.

\bibitem{abani2007unsteady}
N.~Abani and R.~D. Reitz, ``Unsteady turbulent round jets and vortex motion,''
  {\em Physics of Fluids}, vol.~19, no.~12, p.~125102, 2007.

\bibitem{cushman2010turbulent}
B.~Cushman-Roisin, ``Environmental fluid mechanics,'' 2019.

\bibitem{scorer1997dynamics}
R.~S. Scorer and R.~S. Scorer, {\em Dynamics of meteorology and climate}.
\newblock Wiley Chichester, 1997.

\bibitem{han2013characterizations}
Z.~Han, W.~Weng, and Q.~Huang, ``Characterizations of particle size
  distribution of the droplets exhaled by sneeze,'' {\em Journal of The Royal
  Society Interface}, vol.~10, no.~88, p.~20130560, 2013.

\bibitem{Sirignano_2010}
W.~A. Sirignano, {\em Fluid Dynamics and Transport of Droplet and Sprays}.
\newblock Cambridge University Press, 2010.

\bibitem{mansour_2020}
E.~Mansour, R.~Vishinkin, S.~Rihet, W.~Saliba, F.~Fish, P.~Sarfati, and
  H.~Haick, ``Measurement of temperature and relative humidity in exhaled
  breath,'' {\em Sensors and Actuators B: Chemical}, vol.~304, p.~127371, 2020.

\bibitem{abramovich_2003}
G.~Abramovich, {\em The Theory of Turbulent Jets}.
\newblock MIT Press, 2003.

\bibitem{naillon2015evaporation}
A.~Naillon, P.~Duru, M.~Marcoux, and M.~Prat, ``Evaporation with sodium
  chloride crystallization in a capillary tube,'' {\em Journal of Crystal
  Growth}, vol.~422, pp.~52--61, 2015.

\bibitem{derluyn2012salt}
H.~Derluyn, {\em Salt transport and crystallization in porous limestone:
  neutron-X-ray imaging and poromechanical modeling}.
\newblock PhD thesis, ETH Zurich, 2012.

\bibitem{stilianakis2010dynamics}
N.~I. Stilianakis and Y.~Drossinos, ``Dynamics of infectious disease
  transmission by inhalable respiratory droplets,'' {\em Journal of the Royal
  Society Interface}, vol.~7, no.~50, pp.~1355--1366, 2010.

\bibitem{law_2006}
C.~K. Law, {\em Combustion Physics}.
\newblock Cambridge University Press, 2006.

\bibitem{karamouzas2014universal}
I.~Karamouzas, B.~Skinner, and S.~J. Guy, ``Universal power law governing
  pedestrian interactions,'' {\em Physical review letters}, vol.~113, no.~23,
  p.~238701, 2014.

\bibitem{weidmann1993transporttechnik}
U.~Weidmann, ``Transporttechnik der fussg{\"a}nger-transporttechnische
  eigenschaftendes fussg{\"a}ngerverkehrs (literaturstudie),'' {\em Literature
  Research}, vol.~90, 1993.

\bibitem{hsu1994coughing}
J.~Hsu, R.~Stone, R.~Logan-Sinclair, M.~Worsdell, C.~Busst, and K.~Chung,
  ``Coughing frequency in patients with persistent cough: assessment using a 24
  hour ambulatory recorder,'' {\em European Respiratory Journal}, vol.~7,
  no.~7, pp.~1246--1253, 1994.

\bibitem{lloyd2005superspreading}
J.~O. Lloyd-Smith, S.~J. Schreiber, P.~E. Kopp, and W.~M. Getz,
  ``Superspreading and the effect of individual variation on disease
  emergence,'' {\em Nature}, vol.~438, no.~7066, pp.~355--359, 2005.

\bibitem{barbosa2018human}
H.~Barbosa, M.~Barthelemy, G.~Ghoshal, C.~R. James, M.~Lenormand, T.~Louail,
  R.~Menezes, J.~J. Ramasco, F.~Simini, and M.~Tomasini, ``Human mobility:
  Models and applications,'' {\em Physics Reports}, vol.~734, pp.~1--74, 2018.

\bibitem{kolbl2003energy}
R.~K{\"o}lbl and D.~Helbing, ``Energy laws in human travel behaviour,'' {\em
  New Journal of Physics}, vol.~5, no.~1, p.~48, 2003.

\bibitem{duguid1946size}
J.~Duguid, ``The size and the duration of air-carriage of respiratory droplets
  and droplet-nuclei,'' {\em Epidemiology \& Infection}, vol.~44, no.~6,
  pp.~471--479, 1946.

\bibitem{fennelly2020}
K.~P. Fennelly, ``Particle sizes of infectious aerosols: implications for
  infection control,'' {\em The Lancet Respiratory Medicine}, 2020.

\bibitem{chia2020detection}
P.~Y. Chia, K.~K. Coleman, Y.~K. Tan, S.~W.~X. Ong, M.~Gum, S.~K. Lau, X.~F.
  Lim, A.~S. Lim, S.~Sutjipto, P.~H. Lee, {\em et~al.}, ``Detection of air and
  surface contamination by sars-cov-2 in hospital rooms of infected patients,''
  {\em Nature communications}, vol.~11, no.~1, pp.~1--7, 2020.

\bibitem{chen2020short}
W.~Chen, N.~Zhang, J.~Wei, H.-L. Yen, and Y.~Li, ``Short-range airborne route
  dominates exposure of respiratory infection during close contact,'' {\em
  Building and Environment}, p.~106859, 2020.

\bibitem{liu2020reproductive}
Y.~Liu, A.~A. Gayle, A.~Wilder-Smith, and J.~Rockl{\"o}v, ``The reproductive
  number of covid-19 is higher compared to sars coronavirus,'' {\em Journal of
  travel medicine}, 2020.

\bibitem{Leffler2020.05.22.20109231}
C.~T. Leffler, E.~B. Ing, J.~D. Lykins, M.~C. Hogan, C.~A. McKeown, and
  A.~Grzybowski, ``Association of country-wide coronavirus mortality with
  demographics, testing, lockdowns, and public wearing of masks. update july 2,
  2020.,'' {\em medRxiv}, 2020.

\end{thebibliography}
\bibliographystyle{ieeetr}
\end{document}